\documentclass[11pt]{article}

\usepackage[a4paper,margin=1in]{geometry}
\usepackage{amsmath,amssymb,bm}
\usepackage{graphicx}
\usepackage{booktabs}
\usepackage{longtable}
\usepackage{array}
\usepackage{multirow}
\usepackage{caption}
\usepackage{subcaption}
\usepackage{enumitem}
\usepackage{natbib}
\usepackage{xcolor}
\definecolor{changecolor}{RGB}{170,70,0}
\newif\ifshowchanges
\showchangesfalse % control whether changes should be highlighted
\newcommand{\chg}[1]{{\ifshowchanges\color{changecolor}\fi #1}}
\newenvironment{chgblock}{\begingroup\ifshowchanges\color{changecolor}\fi}{\endgroup}
\usepackage[colorlinks=true,linkcolor=blue,citecolor=blue,urlcolor=blue]{hyperref}
\usepackage{microtype}
\usepackage{comment}

\title{A Bayesian Phase I/II basket design with robust information borrowing to identify subtrial-specific optimal biological doses}

\author{
Zhi Cao$^{1*}$, Haiyan Zheng$^{2}$, Pavel Mozgunov$^{1}$\\[0.4em]
\small $^{1}$MRC Biostatistics Unit, University of Cambridge, Cambridge, UK\\
\small $^{2}$Department of Mathematical Sciences, University of Bath, Bath, UK\\
\small $^{*}$zhi.cao@mrc-bsu.cam.ac.uk
}

\date{}

\begin{document}
\maketitle

\begin{abstract}
The objective of modern early oncology dose-finding is to identify an optimal biological dose (OBD), rather than simply the maximum tolerated dose. This change is particularly important for targeted agents and immunotherapies, for which efficacy may plateau before unacceptable toxicity is reached. In basket trials, the dose-toxicity and dose-efficacy relationships may differ across biomarker or disease-defined subtrials, so a single common dose from pooled analysis may be suboptimal. \chg{We propose a flexible exchangeability--non-exchangeability (EXNEX) dose finding design (DF-EXNEX design) for subtrial-specific OBD selection in basket phase I/II trials with binary toxicity and continuous efficacy endpoints.} Patient toxicity is modelled by a monotone logistic regression and efficacy by a flexible quadratic dose-response curve. Robust borrowing is introduced through extended EXNEX mixture priors on the subtrial-specific curve parameters, allowing the strength of borrowing to adapt to the similarity of subtrials. Dose recommendation is based on an admissible set defined by posterior safety and futility rules, and an OBD-oriented utility function combining toxicity and efficacy on comparable scales. The operating characteristics were evaluated in a large-scale simulation study for the basket trial with four subtrials and five dose levels, and 70 scenarios covering all non-redundant combinations of true subtrial-specific OBD locations. \chg{Results showed that, compared with a no-borrowing NEX design, the DF-EXNEX design can increase the correct OBD selection for most scenarios while reducing overly toxic recommendation as final OBD.} The improvement increased with subtrial similarity due to robust information borrowing, but \chg{a small number of mixed low/high OBD scenarios showed negative or near-zero gains}, consistent with occasional over-borrowing towards intermediate doses. These results support robust borrowing for subtrial-specific OBD finding while highlighting the need to monitor borrowing behaviour when true OBDs are widely separated.
\end{abstract}


\section{Introduction}

Dose finding remains a central task in early-phase drug development \citep{Yap2023CONSORTDEFINE, Yan2018PhaseI_II_EffTox}. Historically, many phase I designs were built around the maximum tolerated dose (MTD) motivated by cytotoxic chemotherapy, where a higher dose was often expected to be more efficacious until toxicity became unacceptable. In the era of precision medicine, this assumption is less tenable for many targeted agents, immunotherapies, and biologically active combinations \citep{LeTourneau2009_phaseIreview}. Their efficacy may plateau or even decrease at higher doses, while cumulative or immune-mediated toxicities may continue to increase. Regulatory initiatives such as Project Optimus have therefore reinforced the need for dose optimisation based on the balance of safety, efficacy, pharmacokinetic, pharmacodynamic and other exposure-response evidence rather than on toxicity alone \citep{FDAProjectOptimus2024}. The statistical literature has long recognised that early-phase dose-finding can be formulated as a joint toxicity-efficacy problem, which can be coped with by utility-based designs to select doses according to a clinically specified balance between efficacy and toxicity \citep{ThallCook2004,Lin2020BOIN12,Mandrekar2010}. Such methods are well aligned with the idea of an optimal biological dose (OBD): the clinically acceptable dose with the most favourable benefit-risk profile, especially when dose-efficacy curves are unimodal, non-monotone or plateauing. Therefore, the MTD, even if estimable, may not be the most appropriate recommended dose for further development for targeted agents and immunotherapies.

\chg{Traditionally, early-phase oncology development has often begun with a Phase I dose-escalation study to identify a tolerable dose or recommended Phase II dose, followed by subsequent cohort expansion or parallel studies to evaluate antitumour activity in molecularly or clinically defined subgroups \citep{LeTourneau2009_phaseIreview,FDA2022ExpansionCohorts,Hobbs_basket_review}. This pathway is operationally familiar to practitioners, but the dose selected from all-comers population can be limiting when the best toxicity-efficacy trade-off differs across subgroups. In such cases, an all-comers recommended dose may be inefficient or even clinically inappropriate for some subtrials, motivating an integrated design in which dose finding and subgroup-specific evaluation are conducted within the same basket trial.}

\chg{The dose optimisation problem becomes more complex in basket trials, which evaluate a treatment across several patient subgroups}, often defined by tumour type, molecular alteration, or other biomarkers, within an overarching protocol \citep{Hobbs_basket_review}. \chg{As one type of {\em master protocol} approach \citep{Ouma2022}, basket trials are attractive because they allow efficient evaluation of a treatment across related but potentially heterogeneous subtrials (characterised by the subgroups) to expedite drug development and approval.} For the analysis of basket trials, information borrowing across subtrials can improve estimation and decision-making when treatment effects are similar, but can be harmful when inappropriate borrowing masks heterogeneity \citep{Thall2003,Neuenschwander2016,Jin2020,Ouma2022}. \chg{In dose-finding studies, this issue is especially important because the target is not only whether a treatment works, but which dose should be recommended in each subtrial.} If the true OBD differs across subtrials, a pooled or overly exchangeable analysis may select a wrong dose that is not optimal for any particular subgroup, \chg{whereas a fully no-borrowing analysis is less efficient, although it recognises patient heterogeneity.}

Several recent designs have been proposed for OBD finding in basket or multiple-indication settings. Shotgun-2 was proposed to identify indication-specific OBDs in a Bayesian phase I/II basket design \citep{Chen2023Shotgun2}. ROMI considers dose optimisation for multiple indications using a randomised two-stage basket design and latent cluster borrowing \citep{Wang2024ROMI}. \chg{These related work separated the dose escalation and optimisation stages, while there is still a need for an integral sequential design that uses accumulating toxicity and efficacy outcomes throughout the basket trial to update dose-response models, borrow information robustly across subtrials, recommend the next cohort dose, and finally select subtrial-specific OBDs.} %Therefore,  still remains a need for designs that (i) model both toxicity and efficacy dose-response curves, (ii) select subtrial-specific OBDs, and (iii) borrow information robustly rather than assuming either full exchangeability or full independence.

In this paper, we propose a \chg{dose-finding EXNEX borrowing design (DF-EXNEX design)} for subtrial-specific OBD selection in basket phase I/II trials. The proposed design is a direct dose-finding extension and application of the exchangeability--non-exchangeability modelling idea of \citet{Neuenschwander2016} and the two-dimensional E-BiEXNEX framework for joint toxicity-efficacy evaluation in basket trials \citep{Cao2025EBiEXNEX}. Binary toxicity and continuous efficacy are modelled using dose-response curves within each subtrial. Robust borrowing is introduced through mixture distributions that allow subtrial-specific parameters to be exchangeable, partially exchangeable, or non-exchangeable. The resulting posterior inference is then used to define admissible doses, recommend the next cohort dose, and select a final subtrial-specific OBD. We then compare the \chg{DF-EXNEX design with the NEX/no-borrowing design} in terms of various operating characteristics across comprehensive scenarios.

The remainder of the paper is organised as follows. Section~\ref{sec:methods} introduces the \chg{sequential OBD-finding procedure, endpoint models, and robust EXNEX borrowing}. Section~\ref{sec:simulation} describes prior calibration for desirable properties, simulation scenarios, operating characteristics, and a large-scale simulation study. Section~\ref{sec:results} presents the comparison between the \chg{DF-EXNEX and NEX designs}, including a focused analysis of borrowing-sensitive scenarios. Section~\ref{sec:discussion} discusses implications, limitations, and possible extensions.

\section{Methods}
\label{sec:methods}

\subsection{Trial setting and notation}

Consider a phase I/II dose-finding basket trial with $K$ subtrials and $D$ ordered dose levels $d_1<\cdots<d_D$ for evaluation per subtrial $k=1,\dots, K$. \chg{Let $d_{\mathrm{ref}}$ denote the pre-specified reference dose. The dose-response models below will be written in terms of the log-dose ratio $\log(d_j/d_{\mathrm{ref}})$ to conveniently monitor drug effects at the reference dose.} Patient cohorts will be treated sequentially during the trial. In subtrial $k$, cohort $c$ is treated at dose level $j_{ck}$ and has size $n_{ck}$. For endpoint profiles, the binary toxicity endpoint is intended to represent a dose-limiting toxicity (DLT) or dose-limiting event (DLE) beyond oncology. The continuous efficacy endpoint may represent, for example, tumor size shrinkage \citep{karrison2007}, ctDNA \citep{Bartolomucci2025_ctDNA}, or another quantitative measure of treatment activity. Therefore, we assume the binary toxicity outcome and the continuous efficacy outcome given Common Terminology Criteria for Adverse Events (CTCAE) \citep{CTCAEv4}, and the accumulating recognition of utilising continuous responses in the era of precision medicine. The use of this combination in our model-based phase I/II dose-finding is additionally because it allows toxicity and efficacy to contribute jointly to dose recommendation while retaining a simple and pragmatic dose-curve model.

\subsection{Overview of the sequential OBD-finding design}
\label{sec:sequential-design}

\begin{figure}[!htbp]
\centering
\includegraphics[width=\textwidth]{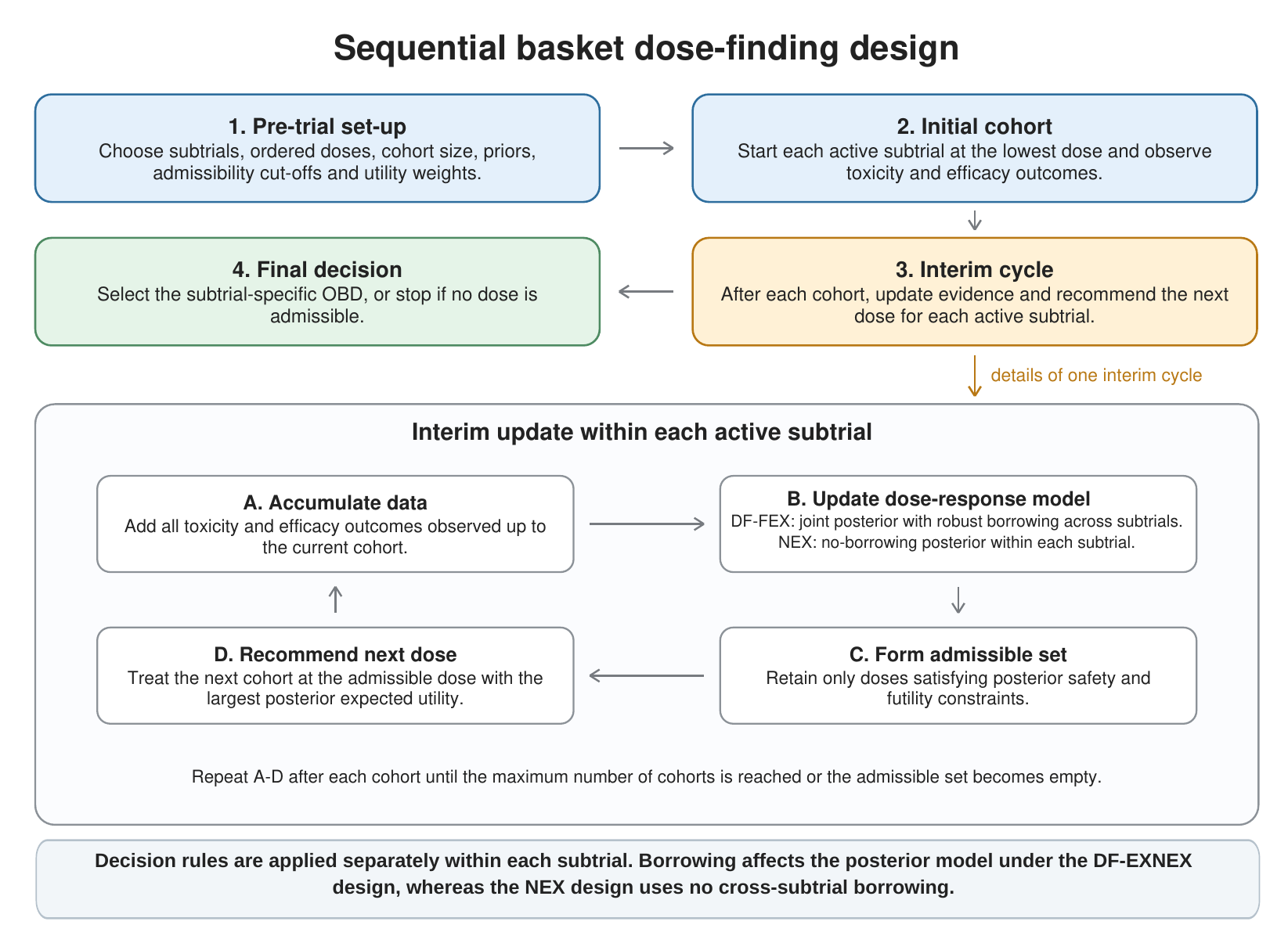}
\caption{Sequential conduct of the basket trial dose-finding design.}
\label{fig:design-flow}
\end{figure}

Figure~\ref{fig:design-flow} summarises the trial conduct into an illustrative flow chart. \chg{Each subtrial starts at the lowest dose and is updated sequentially. At each interim analysis, the accumulated toxicity and efficacy data are used to update the dose-response model, construct the admissible set, and recommend the next dose for each active subtrial.} For ongoing subtrials, the next cohort is assigned to the admissible dose that maximises posterior expected utility, subject to the protocol escalation rule. Formal definitions of the admissible set, utility function, interim recommendation, and final OBD will be given in Section~\ref{ad_utility}. 

\subsection{Dose-toxicity model}

Let $Y_{ck}$ be the number of toxicities observed in cohort $c$ of subtrial $k$. Conditional on the toxicity probability $p_{j_{ck},k}$ at the assigned dose $j$,
\begin{align}
Y_{ck}\mid p_{j,k} &\sim \mathrm{Binomial}(n_{ck},p_{j,k}), \\
\mathrm{logit}\{p_{j,k}\} &= \alpha_k+\exp(\beta_k)\log\left(\frac{d_j}{d_{\mathrm{ref}}}\right),
\qquad j=1,\ldots,D,\quad k=1,\ldots,K .
\end{align}
The exponential transformation of $\beta_k$ enforces monotonicity of toxicity in dose. Define the subtrial-specific toxicity parameter vector
\[
\bm{\theta}^{(T)}_k=(\alpha_k,\beta_k)^\top 
\]
where \(\alpha_k\) controls
the toxicity level at the reference dose and \(\beta_k\) controls the dose-toxicity
slope through the model, for later joint evaluation and robust information borrowing \chg{under Bayesian hierarchical modelling}.

\subsection{Dose-efficacy model}

Let $Z_{ick}$ be the continuous efficacy response for patient $i$ in cohort $c$ of subtrial $k$. Conditional on the mean efficacy $\mu_{j_{ck},k}$ at the assigned dose,
\begin{align}
Z_{ick} &\mid \mu_{j_{ck},k},\sigma_k^2 \sim N(\mu_{j_{ck},k},\sigma_k^2),\\
\mu_{j,k} &= a_k+b_k\log\left(\frac{d_j}{d_{\mathrm{ref}}}\right)+c_k\left\{\log\left(\frac{d_j}{d_{\mathrm{ref}}}\right)\right\}^2 .
\end{align}
where $\sigma_k$ is the unknown standard deviation of patient responses, and the quadratic link function allows various patterns of dose-efficacy curves (e.g. increasing, plateauing, and unimodal). Similarly, we define the subtrial-specific efficacy parameter vector \chg{in the following hierarchical model}:
\[
\bm{\theta}^{(E)}_k=(a_k,b_k,c_k)^\top .
\]
where \(a_k\) is the efficacy level at the
reference dose and \(b_k,c_k\) describe the local linear and quadratic shape of the
dose-efficacy curve.

\subsection{Robust EXNEX borrowing model}
\label{sec:exnex-model}

The proposed borrowing model is motivated by the exchangeable--non-exchangeable (EXNEX, \citet{Neuenschwander2016}) and E-BiEXNEX frameworks \citep{Cao2025EBiEXNEX}, but is adapted here to dose-response modelling. Instead of placing a single high-dimensional joint prior on all toxicity and efficacy curve parameters, we model the dose-toxicity and dose-efficacy curves separately to reduce the model complexity. In early-phase trials, the number of patients is usually limited \citep{Huang2015_early_phase_sample_size}, whereas a fully joint model for all toxicity and efficacy curve parameters would require a large number of covariance and correlation parameters, which can be weakly identified and could make the design difficult to calibrate and implement. We therefore allow robust borrowing within the toxicity curve and within the efficacy curve separately, and then combine the two endpoints to make decisions through the admissibility rules and the specified utility function. This preserves the main purpose of the design, namely subtrial-specific OBD selection, while avoiding an unnecessarily over-parameterised joint model.

For clarity, we introduce latent component indicators
\[
\mathcal{C}^{(T)}_k \in \{1,2,3\}, \qquad \mathcal{C}^{(E)}_k \in \{1,2,3\},
\]
where the three components correspond to full exchangeability, partial exchangeability and full non-exchangeability, respectively. \chg{The first component indicates to borrow information for the whole dose-response curve, the second borrows selected shape parameters while leaving the baseline level non-exchangeable, and the third is the no-borrowing component.}

For the toxicity parameters, define
\[
\omega^{(T)}_h=\Pr\{\mathcal{C}^{(T)}_k=h\},\qquad
\sum_{h=1}^3 \omega^{(T)}_h=1,
\]
and
\[
\bm{\theta}^{(T)}_k \mid \mathcal{C}^{(T)}_k=h,\bm\eta_T \sim F^{(T)}_h,
\qquad h=1,2,3,
\]
where $\bm\eta_T$ \chg{symbolises the toxicity hyperparameters in Equations~\eqref{eq:tox-full-ex},\eqref{eq:tox-partial-ex},\eqref{eq:tox-nex}}. Equivalently, combine over the component indicator,
\begin{equation}
\bm{\theta}^{(T)}_k \mid \bm\eta_T \sim
\omega^{(T)}_1F^{(T)}_1+
\omega^{(T)}_2F^{(T)}_2+
\omega^{(T)}_3F^{(T)}_3 .
\label{eq:tox-mixture-detailed}
\end{equation}
The three toxicity components are defined as bivariate normal distributions:
\begin{align}
F^{(T)}_1 &= N_2\left(\bm{\mu}_T,\bm{\Sigma}_{T,\mathrm{EX}}\right),
\label{eq:tox-full-ex}\\
F^{(T)}_2 &= N_2\left(\bm{\mu}_{T,\mathrm{MIX}},\bm{\Sigma}_{T,\mathrm{MIX}}\right),
\label{eq:tox-partial-ex}\\
F^{(T)}_3 &= N_2\left(\bm{m}_{0T},\bm{\Sigma}_{T,\mathrm{NEX}}\right).
\label{eq:tox-nex}
\end{align}
In the fully exchangeable component $F^{(T)}_1$, $\bm{\mu}_T=(\mu_\alpha,\mu_\beta)^\top$ is the population mean vector shared across subtrials, and
\[
\bm{\Sigma}_{T,\mathrm{EX}}
=
\begin{pmatrix}
\phi_\alpha^2 & \rho_T\phi_\alpha\phi_\beta\\
\rho_T\phi_\alpha\phi_\beta & \phi_\beta^2
\end{pmatrix}.
\]
Here $\phi_\alpha$ and $\phi_\beta$ control the between-subtrial heterogeneity in the shared distribution, and $\rho_T$ is the correlation between the toxicity intercept and slope. Thus, when $\mathcal{C}^{(T)}_k=1$, both $\alpha_k$ and $\beta_k$ are regarded as exchangeable across subtrials. This component borrows information on both the toxicity level at the reference dose and the dose-toxicity slope.

The partially exchangeable component $F^{(T)}_2$ borrows only the toxicity slope. \chg{Its mean vector and covariance matrix are
\[
\bm{\mu}_{T,\mathrm{MIX}}=(m_{0\alpha},\mu_\beta)^\top,\qquad
\bm{\Sigma}_{T,\mathrm{MIX}}=
\begin{pmatrix}
s_{0\alpha}^2 & 0\\
0 & \phi_\beta^2
\end{pmatrix}.
\]
Thus, the toxicity intercept has the fixed operational prior $N(m_{0\alpha},s_{0\alpha}^2)$, whereas the slope is centred on the shared mean $\mu_\beta$ with between-subtrial standard deviation $\phi_\beta$.} This allows subtrials to have different baseline toxicity risks while still sharing information on the dose-toxicity shape.

The fully non-exchangeable component $F^{(T)}_3$ is defined by
\[
\bm{m}_{0T}=(m_{0\alpha},m_{0\beta})^\top,
\qquad
\bm{\Sigma}_{T,\mathrm{NEX}}
=
\begin{pmatrix}
 s_{0\alpha}^2 & \rho_{0T}s_{0\alpha}s_{0\beta}\\
 \rho_{0T}s_{0\alpha}s_{0\beta} & s_{0\beta}^2
\end{pmatrix}.
\]
The quantities $m_{0\alpha},m_{0\beta},s_{0\alpha}$ and $s_{0\beta}$ are fixed before the trial by prior calibration or elicitation from clinical experts. This component acts as an operational prior for a subtrial that should not borrow information from the other subtrials, and corresponds to the no-borrowing stand-alone analysis commonly used as a reference in early-phase basket settings \citep{Cunanan2017_SA_basket_trials,braf-v600}.

The efficacy borrowing model is specified analogously for $\bm{\theta}^{(E)}_k=(a_k,b_k,c_k)^\top$. Define
\[
\omega^{(E)}_h=\Pr\{\mathcal{C}^{(E)}_k=h\},\qquad
\sum_{h=1}^3 \omega^{(E)}_h=1,
\]
and
\[
\bm{\theta}^{(E)}_k \mid \mathcal{C}^{(E)}_k=h,\bm\eta_E \sim F^{(E)}_h,
\qquad h=1,2,3,
\]
where $\bm\eta_E$ \chg{denotes the efficacy hyperparameters in Equations~\ref{eq:eff-full-ex},\ref{eq:eff-partial-ex}, \ref{eq:eff-nex}}. Hence,
\begin{equation}
\bm{\theta}^{(E)}_k \mid \bm\eta_E \sim
\omega^{(E)}_1F^{(E)}_1+
\omega^{(E)}_2F^{(E)}_2+
\omega^{(E)}_3F^{(E)}_3 .
\label{eq:eff-mixture-detailed}
\end{equation}
The three efficacy components are
\begin{align}
F^{(E)}_1 &= N_3\left(\bm{\mu}_E,\bm{\Sigma}_{E,\mathrm{EX}}\right),
\label{eq:eff-full-ex}\\
F^{(E)}_2 &= N_3\left(\bm{\mu}_{E,\mathrm{MIX}},\bm{\Sigma}_{E,\mathrm{MIX}}\right),
\label{eq:eff-partial-ex}\\
F^{(E)}_3 &= N_3\left(\bm{m}_{0E},\bm{\Sigma}_{E,\mathrm{NEX}}\right).
\label{eq:eff-nex}
\end{align}
For the fully exchangeable efficacy component $F^{(E)}_1$, $\bm{\mu}_E=(\mu_a,\mu_b,\mu_c)^\top$ controls the centre of the borrowing distribution for $a_k,b_k,c_k$, and
\[
\bm{\Sigma}_{E,\mathrm{EX}}=\bm{D}_{E,\mathrm{EX}}\bm{R}_{E,\mathrm{EX}}\bm{D}_{E,\mathrm{EX}},
\qquad
\bm{D}_{E,\mathrm{EX}}=\operatorname{diag}(\phi_a,\phi_b,\phi_c),
\]
with
\[
\bm{R}_{E,\mathrm{EX}}
=
\begin{pmatrix}
1 & \rho_{ab} & \rho_{ac}\\
\rho_{ab} & 1 & \rho_{bc}\\
\rho_{ac} & \rho_{bc} & 1
\end{pmatrix}.
\]
This component borrows information on the whole efficacy curve, including the efficacy level at the reference dose, the local linear trend and the curvature.

The partially exchangeable efficacy component $F^{(E)}_2$ borrows only the shape parameters $(b_k,c_k)$. \chg{It is specified using
\[
\bm{\mu}_{E,\mathrm{MIX}}=(m_{0a},\mu_b,\mu_c)^\top,
\]
with block-diagonal covariance
\[
\bm{\Sigma}_{E,\mathrm{MIX}}=
\begin{pmatrix}
s_{0a}^2 & 0 & 0\\
0 & \phi_b^2 & \rho_{bc}\phi_b\phi_c\\
0 & \rho_{bc}\phi_b\phi_c & \phi_c^2
\end{pmatrix}.
\]
Therefore, the reference-dose efficacy level $a_k$ is assigned the fixed operational prior $N(m_{0a},s_{0a}^2)$, while $(b_k,c_k)$ borrow through the shared hyperparameters $(\mu_b,\mu_c)$ and $(\phi_b,\phi_c)$ with correlation $\rho_{bc}$.} This component is useful when subtrials may differ in baseline efficacy level but have similar dose-efficacy shapes.

The fully non-exchangeable efficacy component $F^{(E)}_3$ is defined by
\begin{align*}
\bm{m}_{0E}&=(m_{0a},m_{0b},m_{0c})^\top,\\
\bm{\Sigma}_{E,\mathrm{NEX}}&=\bm{D}_{E,\mathrm{NEX}}\bm{R}_{E,\mathrm{NEX}}\bm{D}_{E,\mathrm{NEX}},\\
\bm{D}_{E,\mathrm{NEX}}&=\operatorname{diag}(s_{0a},s_{0b},s_{0c}).
\end{align*}
where
\[
\bm{R}_{E,\mathrm{NEX}}
=
\begin{pmatrix}
1 & \rho_{0ab} & \rho_{0ac}\\
\rho_{0ab} & 1 & \rho_{0bc}\\
\rho_{0ac} & \rho_{0bc} & 1
\end{pmatrix}.
\]
As in the toxicity model, the mean and scale parameters in the NEX component are fixed operational prior quantities obtained from calibration or historical observation. This component therefore preserves full subtrial heterogeneity and provides robustness against inappropriate borrowing.

For the exchangeable mean parameters, we use weakly informative normal priors centred at calibrated values:
\[
\mu_\alpha \sim N(m_\alpha,v_\alpha^2),\qquad
\mu_\beta \sim N(m_\beta,v_\beta^2),
\]
and
\[
\mu_a \sim N(m_a,v_a^2),\qquad
\mu_b \sim N(m_b,v_b^2),\qquad
\mu_c \sim N(m_c,v_c^2).
\]
The between-subtrial standard deviations are assigned half-normal priors,
\[
\phi_\alpha \sim \mathrm{HN}(z_\alpha^2),\qquad
\phi_\beta \sim \mathrm{HN}(z_\beta^2),
\]
and
\[
\phi_a \sim \mathrm{HN}(z_a^2),\qquad
\phi_b \sim \mathrm{HN}(z_b^2),\qquad
\phi_c \sim \mathrm{HN}(z_c^2),
\]
where $\mathrm{HN}(z^2)$ denotes the positive half of $N(0,z^2)$. Smaller $z$ values imply stronger prior belief that subtrial-specific curve parameters are similar, whereas larger values allow greater between-subtrial heterogeneity. Correlation parameters in the multivariate normal components are assigned uniform priors on $(-1,1)$ subject to the corresponding correlation matrices being positive definite. In the partially exchangeable components, correlations between exchangeable and non-exchangeable blocks are set to zero as a parsimonious default, following the same principle as the E-BiEXNEX formulation.

\subsection{Likelihood and posterior updating}
\begin{chgblock}
Let \(\mathcal{D}_c\) denote all observed toxicity and efficacy data across the
basket by interim analysis \(c\). Specifically, let \(\mathcal{I}_k(c)\) denote
the set of cohorts in subtrial \(k\) whose outcomes have been observed by this
analysis. Conditional on the dose-response parameters, the toxicity likelihood is
\[
L_T(\mathcal{D}_c\mid \bm{\theta}^{(T)})
=
\prod_{k=1}^{K}
\prod_{\ell\in\mathcal{I}_k(c)}
\binom{n_{\ell k}}{Y_{\ell k}}
p_{j_{\ell k},k}^{Y_{\ell k}}
\{1-p_{j_{\ell k},k}\}^{n_{\ell k}-Y_{\ell k}},
\]
where \(j_{\ell k}\) is the dose assigned to cohort \(\ell\) in subtrial \(k\).
The efficacy likelihood is
\[
L_E(\mathcal{D}_c\mid \bm{\theta}^{(E)},\bm{\sigma}^2)
=
\prod_{k=1}^{K}
\prod_{\ell\in\mathcal{I}_k(c)}
\prod_{i=1}^{n_{\ell k}}
(2\pi\sigma_k^2)^{-1/2}
\exp\left[
-\frac{\{Z_{i\ell k}-\mu_{j_{\ell k},k}\}^2}{2\sigma_k^2}
\right].
\]
The joint likelihood is taken as \(L_T L_E\), corresponding to the working
assumption that toxicity and efficacy outcomes are conditionally independent
given their endpoint-specific dose-response parameters.

Let
\[
\bm{\Theta}
=
\left(\bm{\sigma}^2,
\{\mathcal{C}^{(T)}_k,\mathcal{C}^{(E)}_k, \bm{\theta}^{(T)}_k,\bm{\theta}^{(E)}_k\}_{k=1}^K,
\bm{\eta}_T,\bm{\eta}_E
\right),
\]
where \(\bm{\eta}_T\) and \(\bm{\eta}_E\) collect the endpoint-specific
hyperparameters in the DF-EXNEX prior hierarchy. Combining the likelihood with the
EXNEX prior hierarchy gives
\[
\begin{aligned}
	p(\bm{\Theta}\mid\mathcal{D}_c)
	\propto{}&
	L_T(\mathcal{D}_c\mid \bm{\theta}^{(T)})
	L_E(\mathcal{D}_c\mid \bm{\theta}^{(E)},\bm{\sigma}^2)
	p(\bm{\sigma}^2)
	p(\bm{\eta}_T,\bm{\eta}_E)
	\\
	&\times
	\prod_{k=1}^{K}
	p\{\bm{\theta}^{(T)}_k\mid \mathcal{C}^{(T)}_k,\bm{\eta}_T\}
	\Pr(\mathcal{C}^{(T)}_k)
	p\{\bm{\theta}^{(E)}_k\mid \mathcal{C}^{(E)}_k,\bm{\eta}_E\}
	\Pr(\mathcal{C}^{(E)}_k).
\end{aligned}
\]
The posterior is updated after each interim analysis and is used to compute the
posterior admissibility probabilities, posterior expected utilities, interim dose
recommendations and final OBD selections by Markov Chain Monte Carlo (MCMC) samples.
\end{chgblock}

\subsection{Admissible doses and utility}
\label{ad_utility}

At each interim analysis, each subtrial should have an admissible set of doses to guide dose selection under quantified patient safety and futility constraint \citep{ThallCook2004, Lin2020BOIN12}. For subtrial $k$, dose $j$ is admissible if it satisfies both a posterior safety rule and a posterior efficacy rule:
\begin{equation}
\mathcal{A}_k(\mathcal{D}_c)=
\left\{
j \mid
\Pr(p_{j,k}<p_T^\star\mid \mathcal{D}_c)>\epsilon_1,\ 
\Pr(\mu_{j,k}>\mu_E^\star\mid \mathcal{D}_c)>\epsilon_2
\right\}.
\label{eq:admissible}
\end{equation}
Here $p_T^\star$ is the upper acceptable toxicity probability, $\mu_E^\star$ is the lower acceptable mean efficacy, and $\epsilon_1,\epsilon_2$ are posterior probability cut-offs \chg{which quantify the required posterior evidence that a dose is believed to be sufficiently safe and active.} If $\mathcal{A}_k(\mathcal{D}_c)=\varnothing$, subtrial $k$ will be stopped and no OBD is selected for that subtrial.

For any admissible dose $j$ at subtrial $k$, toxicity and efficacy are linked through the weighted utility function combining posterior expected toxicity $p_{j,k}$ and efficacy mean $\mu_{j,k}$:
\begin{equation}
U_k(j)=
\lambda_T\{1-p_{j,k}\}
+
\lambda_E\, g(\mu_{j,k}),
\quad
g(\mu)=\text{logistic}(a_U\mu+b_U)=\frac{1}{1+\exp\{-(a_U\mu+b_U)\}} .
\label{eq:utility}
\end{equation}
The term $1-p_{j,k}$ rewards lower toxicity, while $g(\mu_{j,k})$ maps continuous efficacy onto the same $(0,1)$ scale. In this article, we use a logistic transformation to achieve scale alignment while other functions can be chosen depending on how clinical experts review the endpoints and utility ranges \citep{Pavel2018_benchmark}. The weights $\lambda_T$ and $\lambda_E$ control the relative contribution of safety and efficacy to final utility score. In the simulation study, we set $\lambda_T=\lambda_E=1$ for equal impact of toxicity and transformed efficacy. The constants \(a_U\) and \(b_U\) can be chosen by specifying two clinically
interpretable anchor points. Let \(\mu_L<\mu_U\) denote two efficacy response
values and let \(U^{(E)}_L,U^{(E)}_U\in(0,1)\) denote the corresponding
pre-specified efficacy utility scores, with \(U^{(E)}_L<U^{(E)}_U\). We require
\[
g(\mu_L)=U^{(E)}_L,
\qquad
g(\mu_U)=U^{(E)}_U .
\]
Since
\[
\operatorname{logit}\{g(\mu)\}=a_U\mu+b_U,
\]
the two calibration equations give
\[
a_U=
\frac{
	\operatorname{logit}\{U^{(E)}_U\}
	-
	\operatorname{logit}\{U^{(E)}_L\}
}{
	\mu_U-\mu_L
},
\qquad
b_U=
\operatorname{logit}\{U^{(E)}_L\}-a_U\mu_L .
\]
Thus, \(\mu_L\) and \(\mu_U\) act as clinically meaningful lower and upper
efficacy anchors, while \(U^{(E)}_L\) and \(U^{(E)}_U\) determine how strongly
these efficacy levels contribute to the overall utility.

For illustration in the simulation study, we used \(a_U=10/3\) and
\(b_U=-2.5\). This gives \(g(0)\approx 0.076\) and \(g(1.5)\approx 0.924\),
so efficacy responses near or below zero contribute little to the utility, whereas
responses around 1.5 are already regarded as highly favourable and additional
efficacy gains have diminishing incremental value. Finally, dose selection is driven by the posterior expected utility,
\[
\bar{U}_k(j\mid \mathcal{D}_c)
=
E\{U_k(j)\mid \mathcal{D}_c\},
\]
After interim cohort $c$ of subtrial $k$, if
\(\mathcal{A}_k(\mathcal{D}_c)=\emptyset\), enrolment to subtrial \(k\) is stopped
and no dose is recommended. Otherwise, the next cohort is assigned to the dose
\[
j^{\mathrm{next}}_{k,c+1}
=
\arg\max_{j\in \mathcal{A}_k(\mathcal{D}_c)}
\bar{U}_k(j\mid \mathcal{D}_c).
\]
After the final cohort, the subtrial-specific OBD is
\begin{equation}
	\widehat{d}_k =
	\begin{cases}
		\arg\max_{j\in \mathcal{A}_k(\mathcal{D}_{\mathrm{final}})}
		E\{U_k(j)\mid \mathcal{D}_{\mathrm{final}}\}, & \mathcal{A}_k(\mathcal{D}_{\mathrm{final}})\neq \varnothing,\\
		0, & \mathcal{A}_k(\mathcal{D}_{\mathrm{final}})=\varnothing,
	\end{cases}
	\label{eq:final-obd}
\end{equation}
where $\mathcal{D}_{\mathrm{final}}$ is all patient data collected until the last cohort and $\widehat{d}_k=0$ denotes early stopping or no acceptable OBD.

\begin{comment}
At the final analysis, the subtrial-specific OBD is similarly defined as
\[
\widehat{\mathrm{OBD}}_k
=
\arg\max_{j\in \mathcal{A}_k(\mathcal{D}_{C_{\max}})}
\bar{U}_k(j\mid \mathcal{D}_{C_{\max}}),
\] where $C_{\max}$ is the maximum number of cohorts at each subtrial, provided that the final admissible set is non-empty.
\end{comment}

\section{Simulation study}
\label{sec:simulation}

\subsection{Basic trial settings}

The simulation study considered $K=4$ subtrials and $D=5$ dose levels, with nominal doses $(10,20,30,50,80)$ mg (or other units). The reference dose was dose level 4. According to traditional 3+3 design and common simulation settings \citep{Braun2014PhaseIOncology, Yan2018PhaseI_II_EffTox}, we set each subtrial to enrol cohorts of size 3, with a maximum of 10 cohorts per subtrial. Thus, each subtrial could enrol up to 30 patients. The upper toxicity threshold is $p_T^\star=0.45$ and the lower efficacy threshold is $\mu_E^\star=0$. The calibrated posterior admissibility cut-offs are $\epsilon_1=0.5$ for safety and $\epsilon_2=0.825$ for efficacy (calibration details are in Section \ref{sec:prior-calibration}).

In the simulation study, the \chg{DF-EXNEX design} assigns equal prior weight to the three
borrowing components,
\[
\omega^{(T)}_1=\omega^{(T)}_2=\omega^{(T)}_3=\frac{1}{3},
\qquad
\omega^{(E)}_1=\omega^{(E)}_2=\omega^{(E)}_3=\frac{1}{3}.
\]
The NEX design is obtained by placing all prior mass on the fully non-exchangeable
components:
\[
\omega^{(T)}_3=\omega^{(E)}_3=1,
\qquad
\omega^{(T)}_1=\omega^{(T)}_2=\omega^{(E)}_1=\omega^{(E)}_2=0.
\]
Thus, the \chg{DF-EXNEX design} allows the posterior to adaptively combine full borrowing,
partial borrowing and no borrowing, whereas the NEX design analyses each subtrial
separately without cross-subtrial borrowing. \chg{This comparator keeps the same likelihood, admissibility and utility structure as the DF-EXNEX design, but removes borrowing availability, which represents the conventional no-borrowing approach and preserves full patient heterogeneity across different subtrials.} Additionally, in the simulation study, escalation was limited to at most one dose level above the current dose, while de-escalation to any lower admissible dose was allowed to prevent rapid escalation based on limited early information but permits prompt movement to safer doses when necessary.

The posterior inference is obtained using MCMC with three chains, 20000 iterations per chain and 5000 burn-in iterations. In each scenario, we conducted 5000 simulations to control MCMC standard error for estimating proportion of correct selection (PCS) under 0.7\% \citep{Morris2019}, which is acceptable for our evaluation and will not consume much time. Fixed data generation seed and patient profiles based on pre-generated latent uniform random variables \citep{Pavel2018_benchmark} were used to improve reproducibility across computing environments. Toxicity outcomes were generated by inverse transformation of the uniform variables under the scenario-specific toxicity probabilities, and continuous efficacy outcomes were generated by inverse normal transformation with scenario-specific means and a fixed standard deviation 0.55. This preserves the intended data-generating model while reducing incidental discrepancies across runs and computing environments.

\subsection{Performance metrics}
\label{sec:metrics}

Before comparing the design performances, it is necessary to define all operating characteristics and performance metrics. Let $m\in\{\mathrm{DF-EXNEX},\mathrm{NEX}\}$ denote the design, $s=1,\ldots,S$ for the scenario, $k=1,\ldots,K$ for the subtrial, and $d^\star_{ks}$ the true OBD for scenario $s$ and subtrial $k$. Let $\widehat{d}^{(m,r)}_{ks}$ be the final selected dose in simulation $r=1,\ldots,R$, where $\widehat{d}=0$ denotes early stopping/no OBD. The subtrial-level proportion of correct selection (PCS) is
\begin{equation}
\widehat{\mathrm{PCS}}^{(m)}_{ks}
=
\frac{1}{R}\sum_{r=1}^{R}
I\{\widehat{d}^{(m,r)}_{ks}=d^\star_{ks}\}.
\label{eq:pcs}
\end{equation}

\noindent The scenario-level geometric mean PCS (geom-PCS) is
\begin{equation}
\widehat{\mathrm{\text{geom-}PCS}}^{(m)}_{s}
=
\left\{\prod_{k=1}^{K}
\widehat{\mathrm{PCS}}^{(m)}_{ks}
\right\}^{1/K}
=
\exp\left[
\frac{1}{K}\sum_{k=1}^{K}
\log\{\widehat{\mathrm{PCS}}^{(m)}_{ks}\}
\right].
\label{eq:gpcs}
\end{equation}
The geometric mean was chosen because it penalises imbalanced performance across subtrials, especially sensitive to small quantities. A design that performs very well in some subtrials but poorly in others should not receive the same scenario-level summary as a design that performs consistently across all subtrials. This is appropriate for basket trials where all subtrials are clinically relevant and subtrial-specific OBD selection is the target.

\noindent We define the early-stop rate as follows:
\begin{equation}
\widehat{\mathrm{Drop}}^{(m)}_{ks}
=
\frac{1}{R}\sum_{r=1}^{R}
I\{\widehat{d}^{(m,r)}_{ks}=0\}.
\label{eq:drop}
\end{equation}
The proportion of overly toxic selection (PTS) was evaluated for subtrials in which some high dose is overly toxic but selected as OBD. PTS is an important metric to assess whether the design can expose too many participants to unsafe doses \citep{Wages2021_operating_characteristics}. Generally, PTS can be defined for subtrials as
\begin{equation}
	\widehat{\mathrm{PTS}}^{(m)}_{ks}
	=
	\frac{1}{R}\sum_{r=1}^{R}
	I\{\widehat{d}^{(m,r)}_{ks} \in \{\text{Overly toxic doses}\}\}.
	\label{eq:pts}
\end{equation}

\begin{comment}Specifically, in this simulation study, PTS was defined for subtrials with $d^\star_{ks}\in\{1,2\}$ as
\begin{equation}
\widehat{\mathrm{PTS}}^{(m)}_{ks}
=
\frac{1}{R}\sum_{r=1}^{R}
I\{\widehat{d}^{(m,r)}_{ks}=5\}.
\label{eq:pts}
\end{equation}
\end{comment}
Scenario-level early-stop and PTS summaries were obtained by averaging the corresponding subtrial-level rates over the relevant subtrials. In the simulation study, we also report differences which are computed as DF-EXNEX minus NEX in terms of corresponding metrics, so positive values favour the DF-EXNEX design for PCS and geom-PCS, while negative values favour the DF-EXNEX design for early stopping and PTS.

To quantify the similarity of the true OBD locations within a basket trial for scenario $s$, we define
\begin{equation}
\mathrm{Similarity}_s = \frac{1}{\binom{K}{2}}\sum_{1\leq k<\ell\leq K}
I(d^\star_{ks}=d^\star_{\ell s}).
\label{eq:similarity}
\end{equation}
For $K=4$, this is the proportion of matching true-OBD pairs among the six possible subtrial pairs. The score ranges from 0 (all four true OBDs distinct) to 1 (all four true OBDs identical). In the simulation study, we used this measure as a metric of similarity across subtrials and borrowing strength due to how scenarios were established in Section \ref{sec:scenario_construction}.

\subsection{Prior calibration}
\label{sec:prior-calibration}

Prior calibration was conducted before the large-scale simulation study. This is necessary because Bayesian dose-finding operating characteristics can be sensitive to prior choices, especially under small sample sizes and sequential updating. Inspired by the cyclic calibration algorithm which chooses an operation prior to maximise the posterior estimate of target geom-PCS \citep{Weishi_cyclic_calibration}, we design a sequential procedure to further reduce the computation burden due to the high-dimensional parameter space of the design.

First, priors in NEX-only part ($F_3^{(T)}$ and $F_3^{(E)}$) were calibrated using six one-subtrial scenarios formed by crossing two dose-toxicity curves with three dose-efficacy profiles. These scenarios gave true OBD locations at doses 4, 3, 1, 5, 3, and 2. The aim was to obtain a reasonable independent-analysis baseline across a wide range of OBD locations and endpoint shapes. A broad grid of NEX prior location and scale candidates was explored, preserving substantial prior uncertainty.

Second, the safety and futility cut-offs ($\epsilon_1$ and $\epsilon_2$) were calibrated under the default NEX-only design. Two null scenarios were used: an overly toxic scenario and a futile scenario. The calibration aimed to obtain an early-stop probability (for unacceptable toxicity or futility) of at least 0.75 under these nulls to some tolerance for erroneous early rejection in non-null scenarios. The final calibrated cut-offs were $\epsilon_1=0.5$ for the toxicity rule ($p_T^\star = 0.45$) and $\epsilon_2=0.825$ for the efficacy rule ($\mu_E^\star = 0$).

Third, DF-EXNEX-related prior parameters were calibrated using six scenarios with varying true OBD patterns which introduce different OBD modes and dose curves:
\[
(1,1,1,1),\quad
(4,4,4,4),\quad
(2,3,4,5),\quad
(2,2,2,5),\quad
(4,4,4,2),\quad
(2,2,3,3).
\]
These scenarios include both highly exchangeable and heterogeneous baskets. The DF-EXNEX borrowing-strength parameter candidates ($z_{l}, l\in \{\alpha, \beta, a, b, c\}$) were chosen to span small to large between-subtrial heterogeneity, guided by the robust EXNEX literature \citep{Neuenschwander2016}.

The full calibration details and parameter space are available in Supplementary Material (Section B) for further reference. In the simulation study, the final calibrated prior values were chosen as follows to preserve meaningful uncertainty while providing enough regularisation for sequential posterior updating in small subtrials. For toxicity, the NEX prior location and scale were
\[
\bm m_{0T}=(m_{0\alpha}, m_{0\beta})^\top=(\mathrm{logit}(0.3),-1)^\top,\qquad
\bm s_{0T}=(s_{0\alpha}, s_{0\beta})^\top=(2,2)^\top,
\]
and the DF-EXNEX hyperprior location, scale, and heterogeneity scales were
\begin{align*}
\bm m_T=(m_{\alpha}, m_{\beta})^\top&=(\mathrm{logit}(0.3),-0.5)^\top,\\
\bm s_T=(v_{\alpha}, v_{\beta})^\top&=(1.0,0.5)^\top,\\
\bm z_T=(z_{\alpha}, z_{\beta})^\top&=(0.25,0.25)^\top.
\end{align*}
For efficacy, the NEX prior location and scale were
\[
\bm m_{0E}=(m_{0a}, m_{0b}, m_{0c})^\top=(1.0,0.5,-0.125)^\top,\qquad
\bm s_{0E}=(s_{0a}, s_{0b}, s_{0c})^\top=(0.5,0.25,0.25)^\top,
\]
and the DF-EXNEX hyperprior location, scale, and heterogeneity scales were
\begin{align*}
	\bm m_E=(m_{a}, m_{b}, m_{c})^\top&=(1.2,0.25,0.05)^\top,\\
	\bm s_E=(s_{a}, s_{b}, s_{c})^\top&=(0.25,0.125,0.125)^\top,\\
	\bm z_E=(z_{a}, z_{b}, z_{c})^\top&=(0.25,0.25,0.14)^\top.
\end{align*}

\subsection{Large-scale scenario construction}
\label{sec:scenario_construction}

\begin{figure}[!htbp]
\centering
\includegraphics[width=0.95\textwidth]{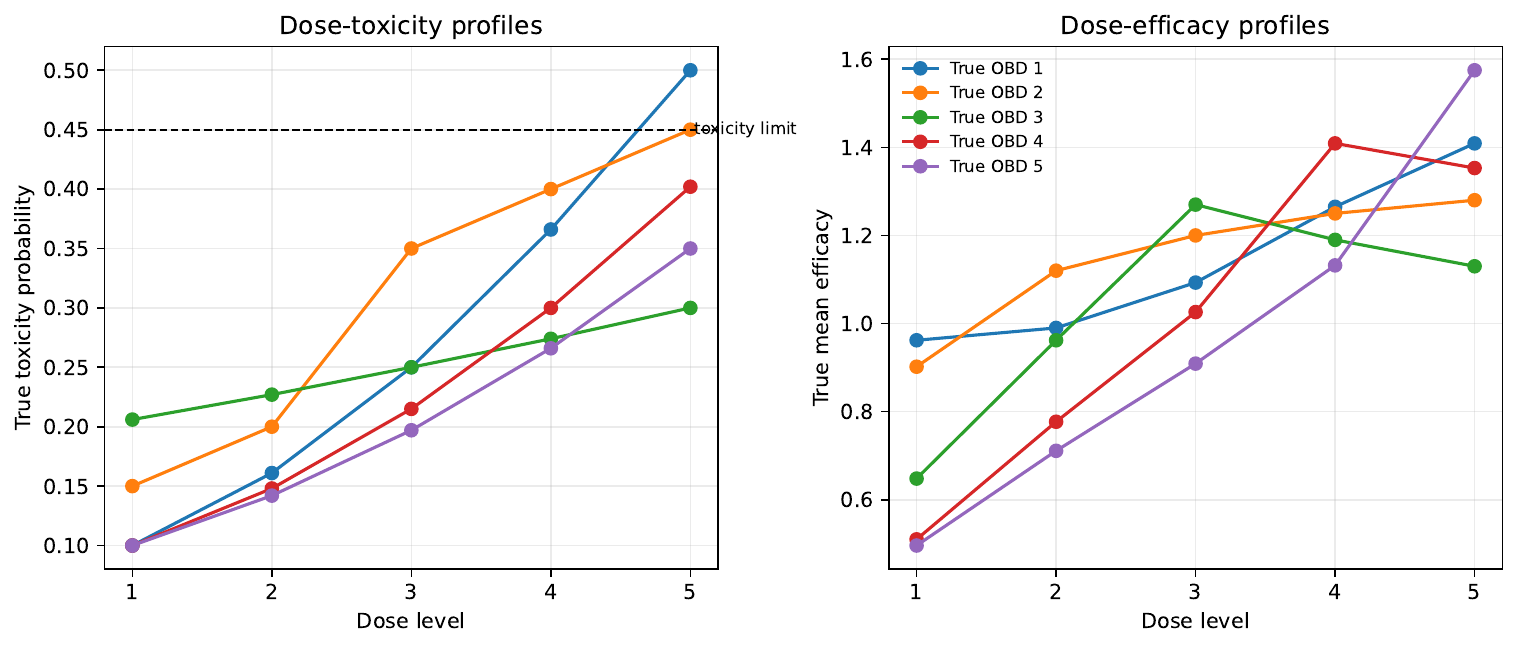}
\caption{Various true dose-toxicity and dose-efficacy profiles used to generate subtrials with true OBD at dose levels 1--5. The horizontal dashed line in the left panel marks the toxicity threshold $p_T^\star=0.45$. Exact value of dose toxicity/efficacy can be found in Supplementary Material (Section A).}
\label{fig:true-profiles}
\end{figure}

The large-scale simulation used five endpoint profiles with various dose-toxicity and dose-efficacy shapes (Figure~\ref{fig:true-profiles}), each designed to have one of the five dose levels as the true OBD location. The scenarios for simulating basket trials were then formed by all non-redundant combinations of four true OBD locations drawn from the set $\{1,2,3,4,5\}$, which gives
$
\binom{5+4-1}{4}=70
$
scenarios. Across the 280 subtrial positions, each true OBD appeared 56 times, giving balanced representation of low, middle, and high true OBDs. Details of true OBDs for each subtrial can be viewed in Supplementary Material (Section C). \chg{We write a scenario by the vector of its four true OBD locations, $(d^\star_{1s},d^\star_{2s},d^\star_{3s},d^\star_{4s})$: for example, $(2,2,5,5)$ denotes two subtrials with true OBD at dose 2 and two subtrials with true OBD at dose 5.}

By mapping the endpoint profiles repetitively to all scenarios formed by different true OBD locations, it is natural to use the similarity score, defined by Equation~\eqref{eq:similarity}, as a measure of borrowing strength since subtrials with same true OBD locations have identical endpoint curves. We aim to provide a uniform metric to present the relationship between operating characteristics and the metric, rather than conducting newly deep research to define a complex similarity measure through dose curves of subtrials which is not the main focus of this paper.

In the simulation study for the NEX design, the no-borrowing structure allowed performance to be evaluated once for each true OBD location and then mapped to the 70 scenarios. For the \chg{DF-EXNEX design}, each of the 70 scenarios was simulated explicitly because borrowing depends on the joint configuration of all subtrials in the basket.

\section{Results}
\label{sec:results}

\subsection{Overall operating characteristics}

Table~\ref{tab:overall-results} summarises the operating characteristics on the average scale across all scenarios. The DF-EXNEX design increased the mean scenario-level geom-PCS across the 70 scenarios, from 0.308 under the NEX design to 0.378, with an absolute gain of 0.070. The mean early-stop rate decreased from 0.057 to 0.039, and the mean PTS among eligible low-OBD scenarios (subtrials with true OBD location at dose 1 or 2 so there exists overly toxic doses in the dosage scheme) decreased from 0.115 to 0.036. The DF-EXNEX design had higher geom-PCS in 67 of the 70 scenarios. And in the same 67 scenarios, it also reduced the mean early-stop rate. Among the 55 PTS-eligible scenarios, the DF-EXNEX design achieved both higher geom-PCS and lower PTS in 52 scenarios.

\begin{table}[!htbp]
\centering
\caption{Overall comparison of the DF-EXNEX and NEX designs across 70 simulated basket trial scenarios.}
\label{tab:overall-results}
\resizebox{\textwidth}{!}{%
\begin{tabular}{lcc}
\toprule
Measure & NEX design & DF-EXNEX\\
\midrule
Mean scenario-level geom-PCS & 0.308 & 0.378 \\
Mean early-stop rate & 0.057 & 0.039 \\
Mean PTS (eligible scenarios) & 0.115 & 0.036 \\
Scenarios with higher geometric mean PCS & 3/70 & 67/70 \\
Scenarios with higher geometric mean PCS and lower early-stop & 3/70 & 67/70 \\
PTS-eligible scenarios with higher geometric mean PCS and lower PTS & 3/55 & 52/55 \\
\bottomrule
\end{tabular}%
}
\end{table}

The summary results show that, generally, robust borrowing improved OBD identification in most scenarios while also reducing undesirable stopping and overly toxic selection in low-OBD settings. The reduction in early stopping suggests that borrowing stabilised posterior estimation when a subtrial alone had limited information. At the same time, the decrease in PTS indicates that borrowing did not simply increase the tendency to recommend higher doses; rather, it improved the ability to identify safe and useful doses under the calibrated prior, utility and admissibility rules.

\subsection{Borrowing and subtrial similarity}

\begin{figure}[!htbp]
\centering
\includegraphics[width=0.85\textwidth]{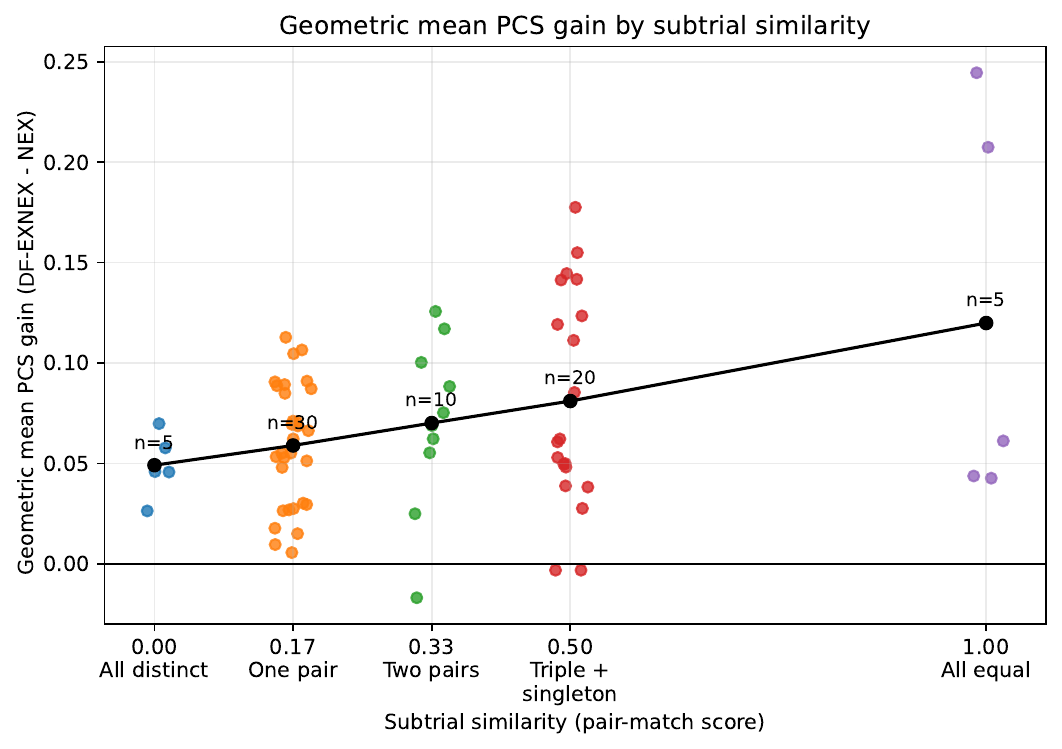}
\caption{Scenario-level geom-PCS gain by subtrial similarity. Each point represents one scenario. The black line joins the mean gain within each similarity class. Similarity is the proportion of matching true-OBD pairs among the six subtrial pairs as defined in Equation~\eqref{eq:similarity}.}
\label{fig:similarity-gain}
\end{figure}

Figure~\ref{fig:similarity-gain} and Table \ref{tab:similarity-summary} shows the geom-PCS gain by subtrial similarity. The average gain increased as more subtrials shared the same true OBD. In all-equal scenarios, the mean gain was 0.120. In all-distinct scenarios, the mean gain was smaller but remained positive at 0.049. The intermediate similarity classes also showed positive average gains: 0.059 for one-pair scenarios, 0.070 for two-pair scenarios, and 0.081 for triple-plus-singleton scenarios. Additionally, Table \ref{tab:similarity-summary} shows that the mean PTS under the DF-EXNEX design also benefit from the robust borrowing: higher subtrial similarity means smaller change to recommend overly toxic doses compared to the NEX design.

\begin{table}[!htbp]
\centering
\caption{Scenario-level performance by subtrial similarity class.}
\label{tab:similarity-summary}
\resizebox{\textwidth}{!}{%
\begin{tabular}{llrrrr}
\toprule
Subtrial similarity & Structure & No. scenarios & Mean geom-PCS gain & Loss scenarios & Mean PTS gain\\
\midrule
0.00 & All distinct & 5 & 0.049 & 0 & -0.075 \\
0.17 & One pair & 30 & 0.059 & 0 & -0.078 \\
0.33 & Two pairs & 10 & 0.070 & 1 & -0.081 \\
0.50 & Triple + singleton & 20 & 0.081 & 2 & -0.082 \\
1.00 & All equal & 5 & 0.120 & 0 & -0.092 \\
\bottomrule
\end{tabular}%
}
\end{table}

This pattern is consistent with the intended behaviour of robust borrowing. When subtrials are similar, the DF-EXNEX design can use information from related subtrials to improve curve estimation and OBD selection. When subtrials are more heterogeneous, the gain is smaller because borrowing should be weaker or less consistently beneficial. Importantly, the all-distinct class still shows positive average gain, suggesting that the robust mixture structure can retain useful borrowing without collapsing to a fully exchangeable analysis.

\chg{To understand when borrowing can be less beneficial, we examined scenarios with negative or small positive scenario-level geom-PCS gain and contrasted them with high-gain scenarios. We regard a scenario as borrowing sensitive when the overall geom-PCS gain is negative or small and at least one subtrial has a PCS loss. This definition is broader than considering only scenarios in which DF-EXNEX loses overall, because a scenario can have a slightly positive average gain while still containing a subtrial in which borrowing pulls selection away from the true OBD.}

\begin{table}[!htbp]
\centering
\caption{Selected low-gain and high-gain scenarios illustrating borrowing-sensitive and borrowing-efficient behaviours. Gains are DF-EXNEX minus NEX.}
\label{tab:borrowing-sensitive-scenarios}
\resizebox{\textwidth}{!}{%
\begin{tabular}{cllrrl}
\toprule
Scenario & Structure & Geom-PCS gain & Min subtrial PCS gain & Borrowing shift\\
\midrule
 $(2,2,5,5)$ & Two pairs & -0.017 & -0.028 &  Low/high groups pulled towards doses 3 and 4\\
 $(2,2,2,5)$ & Triple + singleton & -0.003 & -0.064  & High-OBD singleton pulled down; low group partly shifted upward\\
 $(4,4,4,4)$ & All equal & 0.245 & 0.236  & Compatible subtrials borrow efficiently towards dose 4\\
 $(1,1,1,1)$ & All equal & 0.207 & 0.202 3 & Compatible subtrials borrow efficiently towards dose 1\\
\bottomrule
\end{tabular}%
}
\end{table}

\chg{The two low-gain examples in Table~\ref{tab:borrowing-sensitive-scenarios} illustrate a coherent compromise-dose behaviour in the final OBD selection distribution. In scenario $(2,2,5,5)$, the DF-EXNEX design tended to move the true-OBD-2 subtrials upward towards dose 3 and the true-OBD-5 subtrials downward towards dose 4. Scenario $(2,2,2,5)$, showed a similar pattern: the low-OBD group was partly shifted towards dose 3, while the high-OBD singleton was shifted towards dose 4. These shifts help explain why exact OBD selection can decrease even when early stopping and PTS are reduced. In contrast, the two high-gain all-equal scenarios show the intended benefit of borrowing: when all subtrials are compatible, borrowing stabilises the dose-response estimates and increases correct selection. Detailed posterior dose selection distributions for borrowing sensitive scenarios are provided in the Supplementary Material (Section C).}

\chg{These examples suggest that occasional losses under the DF-EXNEX design are not caused by arbitrary instability. Rather, when low and high OBD subtrials coexist, robust borrowing can produce a reasonable compromise towards intermediate doses. This points to a possible future improvement: borrowing weights or decision rules could be made more explicitly sensitive to evidence of separated OBD clusters, so that strong borrowing is encouraged within compatible subtrial groups but dampened between low and high OBD groups, as explored in related adaptive borrowing work \citep{JackLee_BCHM,Haiyan_discrepancy}.}

\chg{For readers who are interested in model performance by true OBD, the averaged operating characteristics stratified by the true OBD are provided in the Supplementary Material (Section C). In brief, the largest PCS gains occurred when the true OBD was dose 4, while the gain was smaller for true OBD 2 and 5. The DF-EXNEX design nevertheless reduced early stopping at every true OBD and reduced overly toxic selection in the corresponding subtrials.}

\section{Discussion}
\label{sec:discussion}

In this paper, we developed and evaluated a flexible EXNEX dose-finding design for subtrial-specific OBD in basket phase I/II trials. The design combines toxicity and efficacy dose-response modelling, robust borrowing across subtrials, posterior admissibility rules, and a utility-based OBD determination mechanism. In the 70-scenario simulation study covering all non-redundant combinations of true OBD locations across four subtrials, the proposed DF-EXNEX design can improve scenario-level geom-PCS in 67 of 70 scenarios, reduce early stopping in all scenarios, and alleviate overly toxic selection in low-OBD settings.

\chg{The key findings are not simply that borrowing improves average performance, but also that robust borrowing behaves differently across OBD structures.} The subtrial-level geom-PCS gain increased with subtrial similarity as expected, and was largest when all subtrials shared the same true OBD. Additionally, the DF-EXNEX design still improved average performance in (partly) heterogeneous classes, reflecting the robustness of the mixture model. The small number of loss scenarios occurred in mixed low/high OBD patterns, where borrowing can shift selection towards intermediate doses. A clinically interpretable form of over-borrowing is: the design borrows in a direction that is reasonable under partial similarity, but exact OBD selection can suffer when true targets are widely separated.

The use of geom-PCS was helpful because it \chg{summarised performance across subtrials while exposing imbalances}. A simple average PCS can be dominated by large gains in easy or highly borrowable subtrials, whereas the geometric mean penalises a design that performs poorly in one subtrial. \chg{This is aligned with our aim of subtrial-specific OBD selection: in a basket trial, the objective is not merely to obtain a good pooled recommendation, but to recommend an appropriate dose for each patient subgroup.}

However, the proposed design has several limitations. First, dose toxicity was assumed to be logistic shape and efficacy was modelled using a quadratic curve. These are useful and common choices \citep{ThallCook2004}, but may be insufficient for some agents with more complex dose-response patterns. Flexible parametric or semi-parametric efficacy models could be considered \citep{RuitaoLin2023_adaptive_model_selection}. Second, the simulation study used binary toxicity and continuous efficacy endpoints. Other endpoint types such as delayed responses, ordinal responses or time-to-event toxicity may require further model extensions. Third, the borrowing structure was specified at the level of curve parameters rather than directly at the level of OBD locations. This is statistically natural, but the over-borrowing diagnostics suggest that future designs might benefit from borrowing mechanisms that also recognise emerging OBD clusters. Finally, although fixed patient profiles improved reproducibility, the operating characteristics remain specific to the calibrated priors, probability thresholds, cohort size, and scenario set in this paper. \chg{The prior calibration and comprehensive scenario construction are case specific and should be carefully considered in practice.}

We note that some potential extensions of the design exist. \chg{One direction is more sophisticated borrowing, in which the method could identify clusters of similar subtrials \citep{JackLee_BCHM} or enable considerations of pairwise similarity between subtrials} \citep{ Haiyan_discrepancy, Ouma2022}. Another is extending the design to combinations or multi-dimensional dose spaces, where the OBD is a dose pair or regimen rather than a single selection \citep{Weishi_dosecombo, Libby2025_dosecomboP1}. Finally, a more combined decision framework can be developed to incorporate patient-level benefit-risk preferences, posterior uncertainty guidance, and explicit opinions from clinical experts \citep{Thall2024BayesianPrecisionMedicine}.

Overall, the DF-EXNEX design provides a useful and robust framework for OBD selection in basket dose-finding trials. It improves performance in most scenarios and has clear advantages when subtrials are similar. At the same time, the analysis of borrowing-sensitive scenarios highlights the need to diagnose and control possible over-borrowing when true OBDs are widely separated. This balance between efficient information sharing and protection against heterogeneity could be important to subtrial-specific dose optimisation in precision oncology.

%\section*{Code availability}

\section*{Acknowledgements}

ZC receives the doctoral training scholarship for PhD in Biostatistics by the MRC Trials Methodology Research Partnership Doctoral Training Partnership (Grant Ref: MR/W006049/1). This report is an independent research supported by the National Institute for Health and Care Research (NIHR300576). The views expressed in this publication are those of the authors and not necessarily those of the NHS, the National Institute for Health and Care Research or the Department of Health and Social Care. PM also received funding from the UK Medical Research Council (\chg{MC\_UU\_00040/3}). Dr Zheng's contribution to this manuscript was supported by Cancer Research UK (RCCCDF-May24/100001).

\section*{Declaration of conflicting interests}

The authors declare no financial conflict of interest relevant to this manuscript.

\bibliographystyle{plainnat}
\bibliography{references}

\end{document}

\maketitle

\appendix

\section{Endpoint profiles used in the simulation study}

Table~S1 gives the five dose-toxicity and dose-efficacy profiles used to generate true OBD locations at doses 1--5. These profiles were combined to form the 70 large-scale basket scenarios.

\begin{table}[!htbp]
\centering
\caption{Endpoint profiles used to generate the five true OBD locations. Entries are toxicity probabilities followed by mean efficacy responses at doses 1--5.}
\label{tab:true-endpoint-profiles}
\begin{tabular}{clrrrrr}
\toprule
True OBD & Endpoint & Dose 1 & Dose 2 & Dose 3 & Dose 4 & Dose 5\\
\midrule
1 & Toxicity & 0.100 & 0.161 & 0.250 & 0.366 & 0.500 \\
1 & Efficacy & 0.962 & 0.990 & 1.093 & 1.265 & 1.409 \\
2 & Toxicity & 0.150 & 0.200 & 0.350 & 0.400 & 0.450 \\
2 & Efficacy & 0.902 & 1.120 & 1.200 & 1.250 & 1.280 \\
3 & Toxicity & 0.206 & 0.227 & 0.250 & 0.274 & 0.300 \\
3 & Efficacy & 0.648 & 0.962 & 1.270 & 1.190 & 1.130 \\
4 & Toxicity & 0.100 & 0.148 & 0.215 & 0.300 & 0.402 \\
4 & Efficacy & 0.510 & 0.777 & 1.026 & 1.409 & 1.353 \\
5 & Toxicity & 0.100 & 0.142 & 0.197 & 0.266 & 0.350 \\
5 & Efficacy & 0.496 & 0.711 & 0.909 & 1.132 & 1.575 \\
\bottomrule
\end{tabular}
\end{table}

\section{Prior calibration details and parameter spaces}
\label{supp:prior-calibration-details}

This section gives the calibration scenarios, target criteria and candidate parameter spaces used to specify the operational priors and hyperpriors in the simulation study. The calibration was carried out before the large-scale simulation study which aimed not to tune the design to a single favourable setting, but to obtain a stable operating prior across a range of dose-toxicity and dose-efficacy shapes. This is important because Bayesian dose-finding designs may be sensitive to prior choices when the number of patients per subtrial is small and posterior updating is repeated after each cohort.

The calibration proceeded in three steps. First, the non-exchangeable operational priors were calibrated in one-subtrial scenarios under the NEX design (no cross-subtrial borrowing). Second, the safety and futility probability cut-offs defining the admissible set were calibrated under null scenarios. Third, the exchangeable components of the DF-EXNEX design were calibrated in multi-subtrial basket scenarios with varying degrees of between-subtrial similarity.

\subsection{Calibration of NEX operational priors}
\label{supp:nex-prior-calibration}

The first step calibrated the NEX components, denoted by \(F^{(T)}_3\) and \(F^{(E)}_3\) in the main text. For toxicity, the operational prior was written as
\[
\alpha_k \sim N(m_{0\alpha},s_{0\alpha}^2),
\qquad
\beta_k \sim N(m_{0\beta},s_{0\beta}^2),
\]
and for efficacy as
\[
a_k \sim N(m_{0a},s_{0a}^2),
\qquad
b_k \sim N(m_{0b},s_{0b}^2),
\qquad
c_k \sim N(m_{0c},s_{0c}^2).
\]
These parameters define the no-borrowing analysis baseline and are deliberately allowed to retain substantial prior uncertainty.

The dose levels used in calibration were the same as in the simulation study:
\[
(10,20,30,50,80),
\]
and six one-subtrial calibration scenarios were formed by crossing two dose-toxicity profiles with three dose-efficacy profiles. The toxicity profiles were
\begin{align*}
	T_1 &= (0.10,0.20,0.30,0.40,0.50),\\
	T_2 &= (0.10,0.15,0.20,0.30,0.40),
\end{align*}
and the efficacy profiles were
\begin{align*}
	E_1 &= (0.15,0.75,1.05,1.25,1.65),\\
	E_2 &= (0.80,1.05,1.45,1.25,1.05),\\
	E_3 &= (1.20,1.40,1.55,1.56,1.58).
\end{align*}
The resulting six calibration scenarios are shown in Table~\ref{tab:supp-nex-calibration-scenarios}. The first scenario has its largest true utility at dose 5 before imposing admissibility; however, dose 5 exceeds the toxicity threshold of 0.45, so the true OBD is dose 4.

\begin{table}[!htbp]
	\centering
	\caption{One-subtrial scenarios used for NEX operational-prior calibration. The true OBD is defined after applying the toxicity and efficacy admissibility constraints.}
	\label{tab:supp-nex-calibration-scenarios}
	\begin{tabular}{cccc}
		\toprule
		Calibration scenario & Toxicity profile & Efficacy profile & True OBD dose \\
		\midrule
		1 & \(T_1\) & \(E_1\) & 4 \\
		2 & \(T_1\) & \(E_2\) & 3 \\
		3 & \(T_1\) & \(E_3\) & 1 \\
		4 & \(T_2\) & \(E_1\) & 5 \\
		5 & \(T_2\) & \(E_2\) & 3 \\
		6 & \(T_2\) & \(E_3\) & 2 \\
		\bottomrule
	\end{tabular}
\end{table}

The NEX candidate parameter space is given in Table~\ref{tab:supp-nex-parameter-space}. The grid spans a broad set of prior centres and prior standard deviations for the toxicity and efficacy curve parameters. Because the calibration was embedded within a sequential dose-finding simulation and repeated over many candidates, 200 MCMC iterations were used during this calibration stage to reduce computational burden. The selected prior was then conducted forward to subsequent calibration steps and the large-scale simulation study.

\begin{table}[!htbp]
	\centering
	\caption{Candidate parameter space for calibration of the NEX operational priors. The notation \(\operatorname{logit}(A)\) means that the logit transformation is applied to each element in the set \(A\).}
	\label{tab:supp-nex-parameter-space}
	\begin{tabular}{llll}
		\toprule
		Endpoint & Parameter & Prior quantity & Candidate values \\
		\midrule
		Toxicity & \(\alpha_k\) & \(m_{0\alpha}\) & \(\operatorname{logit}\{0.1,0.2,0.3,0.4,0.5\}\) \\
		Toxicity & \(\alpha_k\) & \(s_{0\alpha}\) & \(\{1,1.5,2\}\) \\
		Toxicity & \(\beta_k\) & \(m_{0\beta}\) & \(\{-0.5,-0.25,0,0.25,0.5\}\) \\
		Toxicity & \(\beta_k\) & \(s_{0\beta}\) & \(\{0.5,1,2\}\) \\
		\midrule
		Efficacy & \(a_k\) & \(m_{0a}\) & \(\{-0.25,0,0.5,0.75,1,1.25\}\) \\
		Efficacy & \(a_k\) & \(s_{0a}\) & \(\{0.5,1,2\}\) \\
		Efficacy & \(b_k\) & \(m_{0b}\) & \(\{-0.5,-0.25,0,0.25,0.5\}\) \\
		Efficacy & \(b_k\) & \(s_{0b}\) & \(\{0.5,1,2\}\) \\
		Efficacy & \(c_k\) & \(m_{0c}\) & \(\{-0.5,-0.25,0,0.25,0.5\}\) \\
		Efficacy & \(c_k\) & \(s_{0c}\) & \(\{0.5,1,2\}\) \\
		\bottomrule
	\end{tabular}
\end{table}

\subsection{Calibration of safety and futility cut-offs}
\label{supp:cutoff-calibration}

The second step calibrated the posterior probability cut-offs used to define the admissible set. For subtrial \(k\), after observing data \(\mathcal D_c\) up to cohort \(c\), a dose \(j\) was considered admissible if it satisfied both
\[
\Pr\{p_{j,k}<p_{\max}\mid \mathcal D_c\}\geq \epsilon_1,
\qquad
\Pr\{\mu_{j,k}>\mu_{\min}\mid \mathcal D_c\}\geq \epsilon_2,
\]
where \(p_{\max}=0.45\) is the maximum acceptable toxicity probability and \(\mu_{\min}=0\) is the minimum acceptable efficacy response. If no dose satisfied both conditions, the subtrial was stopped and no OBD was returned.

The cut-offs \(\epsilon_1\) and \(\epsilon_2\) were calibrated under the default NEX design, since this is the no-borrowing baseline which is common in clinical trials. Two null scenarios were used. The first was an overly toxic scenario, with \(p_j=0.45\) and \(\mu_j=2.00\) for all doses \(j=1,\ldots,5\). The second was a futile scenario, with \(p_j=0.05\) and \(\mu_j=0\) for all doses. These settings isolate the two reasons why early stopping should occur: unacceptable toxicity and insufficient efficacy.

The candidate grid was
\[
\epsilon_1,\epsilon_2 \in \{0.200,0.225,0.250,\ldots,1.000\}.
\]
For a given null scenario \(r\), let \(\mathrm{Stop}_{r}\) denote the event that the design stops without selecting an OBD. The calibration target was
\[
\Pr_r(\mathrm{Stop}_{r})\geq 0.75
\]
for both null scenarios. This target was chosen to give a high probability of stopping when the treatment is truly unsuitable, while retaining some tolerance against erroneous early rejection in non-null settings. The final calibrated values were
\[
\epsilon_1=0.5,
\qquad
\epsilon_2=0.825.
\]

\subsection{Calibration of DF-EXNEX-related hyperpriors}
\label{supp:ex-prior-calibration}

The third step calibrated the DF-EXNEX-related hyperpriors and heterogeneity scales used in the DF-EXNEX design. This was done after fixing the NEX operational prior and the admissibility cut-offs. Six four-subtrial basket scenarios were used:
\[
(1,1,1,1),\quad
(4,4,4,4),\quad
(2,3,4,5),\quad
(2,2,2,5),\quad
(4,4,4,2),\quad
(2,2,3,3),
\]
where each number gives the true OBD location in one subtrial. These scenarios include baskets in which all subtrials share the same true OBD, baskets with a single discordant subtrial, and baskets with more heterogeneous OBD patterns. They were therefore used to evaluate both beneficial borrowing and robustness against inappropriate borrowing.

The five OBD-generating dose-toxicity and dose-efficacy profiles are shown in Table~\ref{tab:supp-obd-profiles}. Profile \(d_m\) denotes a curve pair for which dose \(m\) is the true OBD after applying the admissibility rules and utility definition.

\begin{table}[!htbp]
	\centering
	\caption{Dose-toxicity and dose-efficacy profiles used to generate DF-EXNEX calibration scenarios and the large-scale simulation scenarios. Profile \(d_m\) has true OBD dose \(m\).}
	\label{tab:supp-obd-profiles}
	\begin{tabular}{cll}
		\toprule
		Profile & Toxicity probabilities at doses 1--5 & Efficacy means at doses 1--5 \\
		\midrule
		\(d_1\) & \((0.100,0.161,0.250,0.366,0.500)\) & \((0.962,0.990,1.093,1.265,1.409)\) \\
		\(d_2\) & \((0.150,0.200,0.350,0.400,0.450)\) & \((0.902,1.120,1.200,1.250,1.280)\) \\
		\(d_3\) & \((0.206,0.227,0.250,0.274,0.300)\) & \((0.648,0.962,1.270,1.190,1.130)\) \\
		\(d_4\) & \((0.100,0.148,0.215,0.300,0.402)\) & \((0.510,0.777,1.026,1.409,1.353)\) \\
		\(d_5\) & \((0.100,0.142,0.197,0.266,0.350)\) & \((0.496,0.711,0.909,1.132,1.575)\) \\
		\bottomrule
	\end{tabular}
\end{table}

The DF-EXNEX calibration involved three types of candidate quantities: hyperprior centres, hyperprior standard deviations, and between-subtrial heterogeneity scales. For toxicity, the exchangeable mean parameters satisfy
\[
\mu_\alpha \sim N(m_\alpha,v_\alpha^2),
\qquad
\mu_\beta \sim N(m_\beta,v_\beta^2),
\]
and the between-subtrial standard deviations satisfy
\[
\phi_\alpha \sim \mathrm{HN}(z_\alpha^2),
\qquad
\phi_\beta \sim \mathrm{HN}(z_\beta^2).
\]
For efficacy,
\[
\mu_a \sim N(m_a,v_a^2),
\qquad
\mu_b \sim N(m_b,v_b^2),
\qquad
\mu_c \sim N(m_c,v_c^2),
\]
and
\[
\phi_a \sim \mathrm{HN}(z_a^2),
\qquad
\phi_b \sim \mathrm{HN}(z_b^2),
\qquad
\phi_c \sim \mathrm{HN}(z_c^2).
\]
Here \(\mathrm{HN}(z^2)\) denotes a half-normal distribution with scale \(z\). Smaller \(z\) values imply stronger prior belief that subtrial-specific parameters are close to one another, whereas larger values allow greater heterogeneity. The candidate values for the heterogeneity scales were chosen to span small to large between-subtrial heterogeneity, following the borrowing-strength interpretation used in robust EXNEX designs.

\begin{table}[!htbp]
	\centering
	\caption{Candidate values for DF-EXNEX heterogeneity scales. These parameters control the prior degree of borrowing across subtrials.}
	\label{tab:supp-ex-borrowing-strength}
	\begin{tabular}{lll}
		\toprule
		Endpoint & Heterogeneity scale & Candidate values \\
		\midrule
		Toxicity & \(z_\alpha\) & \(\{0.125,0.25,0.5\}\) \\
		Toxicity & \(z_\beta\) & \(\{0.125,0.25,0.5\}\) \\
		Efficacy & \(z_a\) & \(\{0.125,0.25,0.5\}\) \\
		Efficacy & \(z_b\) & \(\{0.125,0.25,0.5\}\) \\
		Efficacy & \(z_c\) & \(\{0.07,0.14,0.28\}\) \\
		\bottomrule
	\end{tabular}
\end{table}

\begin{table}[!htbp]
	\centering
	\caption{Candidate values for DF-EXNEX hyperprior centres.}
	\label{tab:supp-ex-centres}
	\resizebox{\textwidth}{!}{%
	\begin{tabular}{lll}
		\toprule
		Endpoint & Hyperprior centre & Candidate values \\
		\midrule
		Toxicity & \(m_\alpha\) & \(\operatorname{logit}\{0.1,0.2,0.3,0.4,0.5\}\) \\
		Toxicity & \(m_\beta\) & \(\{-1,-0.75,-0.5,-0.25,0,0.25,0.5,0.75,1\}\) \\
		Efficacy & \(m_a\) & \(\{0.5,0.6,0.7,0.8,0.9,1.0,1.1,1.2,1.3,1.4\}\) \\
		Efficacy & \(m_b\) & \(\{-0.5,-0.25,0,0.25,0.5\}\) \\
		Efficacy & \(m_c\) & \(\{-0.25,-0.20,-0.15,-0.10,-0.05,0,0.05,0.10,0.15,0.20,0.25\}\) \\
		\bottomrule
	\end{tabular}%
	}
\end{table}

\begin{table}[!htbp]
	\centering
	\caption{Candidate values for DF-EXNEX hyperprior standard deviations. These quantities determine how concentrated the prior on the exchangeable mean parameters is around its calibrated centre.}
	\label{tab:supp-ex-scales}
	\begin{tabular}{lll}
		\toprule
		Endpoint & Hyperprior standard deviation & Candidate values \\
		\midrule
		Toxicity & \(v_\alpha\) & \(\{0.5,1,1.5,2\}\) \\
		Toxicity & \(v_\beta\) & \(\{0.25,0.5,0.75,1\}\) \\
		Efficacy & \(v_a\) & \(\{0.125,0.25,0.5,1\}\) \\
		Efficacy & \(v_b\) & \(\{0.125,0.25,0.5,1\}\) \\
		Efficacy & \(v_c\) & \(\{0.125,0.25,0.5,1\}\) \\
		\bottomrule
	\end{tabular}
\end{table}

Table~\ref{tab:supp-ex-borrowing-strength},\ref{tab:supp-ex-centres},\ref{tab:supp-ex-scales} show the prior candidates used for DF-EXNEX design, we then applied the cyclic calibration procedure over the paramter space to identify the prior which maximises mean geom-PCS across all tested scenarios.

\subsection{Final calibrated prior values}
\label{supp:final-calibrated-priors}

The final calibrated prior values used in the simulation study are summarised below. For the toxicity NEX component,
\[
\bm m_{0T}=(m_{0\alpha},m_{0\beta})^\top
=\{\operatorname{logit}(0.3),-1\}^\top,
\qquad
\bm s_{0T}=(s_{0\alpha},s_{0\beta})^\top=(2,2)^\top.
\]
For the toxicity DF-EXNEX hyperpriors,
\begin{align*}
\bm m_T=(m_\alpha,m_\beta)^\top&=\{\operatorname{logit}(0.3),-0.5\}^\top,\\
\bm v_T=(v_\alpha,v_\beta)^\top&=(1.0,0.5)^\top,\\
\bm z_T=(z_\alpha,z_\beta)^\top&=(0.25,0.25)^\top.
\end{align*}
For the efficacy NEX component,
\[
\bm m_{0E}=(m_{0a},m_{0b},m_{0c})^\top=(1.0,0.5,-0.125)^\top,
\qquad
\bm s_{0E}=(s_{0a},s_{0b},s_{0c})^\top=(0.5,0.25,0.25)^\top.
\]
For the efficacy DF-EXNEX hyperpriors,
\begin{align*}
	\bm m_E=(m_a,m_b,m_c)^\top&=(1.2,0.25,0.05)^\top,\\
	\bm v_E=(v_a,v_b,v_c)^\top&=(0.25,0.125,0.125)^\top,\\
	\bm z_E=(z_a,z_b,z_c)^\top&=(0.25,0.25,0.14)^\top.
\end{align*}
These values were selected to retain meaningful prior uncertainty, while providing enough regularisation for sequential posterior updating in small subtrials.

\section{Supplementary results}

\subsection{Full scenario-level results}

Table~\ref{tab:full-scenario-results} provides the scenario-level operating characteristics for all 70 scenarios.

\begin{landscape}
\begin{longtable}{rllrrrrr}
\caption{Scenario-level operating characteristics for all 70 scenarios. "Gain" is the difference of geom-PCS (DF-EXNEX minus NEX), "Drop gain" is the difference of early-stop rate. For geom-PCS, positive gain favours DF-EXNEX, while for "Drop gain" and PTS negative value prefers it.}\label{tab:full-scenario-results}\\
\toprule
Scenario & Pattern & Structure & DF-EXNEX geom-PCS & NEX geom-PCS & Gain & Drop gain & PTS gain\\
\midrule
\endfirsthead
\toprule
Scenario & Pattern & Structure & DF-EXNEX geom-PCS & NEX geom-PCS & Gain & Drop gain & PTS gain\\
\midrule
\endhead
1 & 1-2-3-4 & All distinct & 0.341 & 0.271 & 0.070 & -0.019 & -0.082 \\
2 & 1-2-4-5 & All distinct & 0.369 & 0.323 & 0.046 & -0.012 & -0.073 \\
3 & 1-3-3-5 & One pair & 0.310 & 0.280 & 0.030 & -0.025 & -0.060 \\
4 & 1-3-5-5 & One pair & 0.360 & 0.333 & 0.026 & -0.017 & -0.048 \\
5 & 2-3-4-4 & One pair & 0.415 & 0.302 & 0.113 & -0.019 & -0.094 \\
6 & 1-2-3-3 & One pair & 0.296 & 0.245 & 0.051 & -0.029 & -0.081 \\
7 & 1-2-4-4 & One pair & 0.391 & 0.301 & 0.091 & -0.013 & -0.084 \\
8 & 1-1-2-2 & Two pairs & 0.329 & 0.254 & 0.075 & -0.016 & -0.092 \\
9 & 1-3-4-5 & All distinct & 0.368 & 0.310 & 0.058 & -0.016 & -0.060 \\
10 & 1-1-2-3 & One pair & 0.313 & 0.244 & 0.069 & -0.023 & -0.082 \\
11 & 1-2-3-5 & All distinct & 0.318 & 0.292 & 0.026 & -0.020 & -0.072 \\
12 & 2-3-3-4 & One pair & 0.357 & 0.272 & 0.085 & -0.026 & -0.092 \\
13 & 1-4-5-5 & One pair & 0.422 & 0.369 & 0.053 & -0.010 & -0.052 \\
14 & 2-2-3-3 & Two pairs & 0.311 & 0.256 & 0.055 & -0.032 & -0.095 \\
15 & 1-2-2-4 & One pair & 0.345 & 0.283 & 0.062 & -0.016 & -0.092 \\
16 & 1-1-1-1 & All equal & 0.439 & 0.232 & 0.207 & -0.014 & -0.083 \\
17 & 1-1-4-5 & One pair & 0.396 & 0.309 & 0.087 & -0.012 & -0.068 \\
18 & 2-5-5-5 & Triple + singleton & 0.412 & 0.415 & -0.003 & -0.013 & -0.068 \\
19 & 2-2-3-5 & One pair & 0.320 & 0.305 & 0.015 & -0.023 & -0.087 \\
20 & 2-2-4-4 & Two pairs & 0.403 & 0.315 & 0.088 & -0.016 & -0.095 \\
21 & 4-4-5-5 & Two pairs & 0.511 & 0.411 & 0.100 & -0.010 & -- \\
22 & 5-5-5-5 & All equal & 0.519 & 0.475 & 0.044 & -0.010 & -- \\
23 & 3-3-3-4 & Triple + singleton & 0.347 & 0.262 & 0.085 & -0.031 & -- \\
24 & 3-3-5-5 & Two pairs & 0.360 & 0.335 & 0.025 & -0.023 & -- \\
25 & 1-1-2-5 & One pair & 0.344 & 0.290 & 0.053 & -0.015 & -0.075 \\
26 & 2-3-4-5 & All distinct & 0.370 & 0.325 & 0.046 & -0.018 & -0.086 \\
27 & 2-3-5-5 & One pair & 0.354 & 0.349 & 0.006 & -0.017 & -0.079 \\
28 & 3-3-4-5 & One pair & 0.378 & 0.312 & 0.066 & -0.024 & -- \\
29 & 2-2-3-4 & One pair & 0.352 & 0.284 & 0.069 & -0.023 & -0.095 \\
30 & 3-4-4-5 & One pair & 0.450 & 0.345 & 0.105 & -0.017 & -- \\
31 & 2-4-4-5 & One pair & 0.431 & 0.360 & 0.071 & -0.013 & -0.087 \\
32 & 1-1-1-4 & Triple + singleton & 0.400 & 0.258 & 0.141 & -0.013 & -0.079 \\
33 & 1-2-5-5 & One pair & 0.357 & 0.347 & 0.010 & -0.013 & -0.065 \\
34 & 3-4-5-5 & One pair & 0.426 & 0.371 & 0.055 & -0.018 & -- \\
35 & 4-4-4-5 & Triple + singleton & 0.538 & 0.383 & 0.155 & -0.009 & -- \\
36 & 1-2-2-3 & One pair & 0.303 & 0.255 & 0.048 & -0.024 & -0.090 \\
37 & 3-3-3-5 & Triple + singleton & 0.320 & 0.281 & 0.039 & -0.031 & -- \\
38 & 1-2-2-5 & One pair & 0.321 & 0.304 & 0.018 & -0.016 & -0.082 \\
39 & 3-5-5-5 & Triple + singleton & 0.426 & 0.399 & 0.028 & -0.017 & -- \\
40 & 1-2-2-2 & Triple + singleton & 0.319 & 0.266 & 0.053 & -0.019 & -0.097 \\
41 & 1-1-3-4 & One pair & 0.349 & 0.259 & 0.089 & -0.018 & -0.074 \\
42 & 1-4-4-5 & One pair & 0.435 & 0.344 & 0.091 & -0.010 & -0.067 \\
43 & 2-4-5-5 & One pair & 0.414 & 0.387 & 0.027 & -0.012 & -0.079 \\
44 & 1-1-4-4 & Two pairs & 0.404 & 0.287 & 0.117 & -0.013 & -0.077 \\
45 & 2-2-2-2 & All equal & 0.321 & 0.278 & 0.043 & -0.020 & -0.100 \\
46 & 1-3-4-4 & One pair & 0.395 & 0.289 & 0.106 & -0.016 & -0.075 \\
47 & 2-2-5-5 & Two pairs & 0.346 & 0.363 & -0.017 & -0.014 & -0.076 \\
48 & 1-1-3-3 & Two pairs & 0.296 & 0.234 & 0.062 & -0.028 & -0.070 \\
49 & 1-1-5-5 & Two pairs & 0.401 & 0.332 & 0.069 & -0.012 & -0.058 \\
50 & 2-3-3-5 & One pair & 0.322 & 0.293 & 0.030 & -0.026 & -0.085 \\
51 & 2-2-2-3 & Triple + singleton & 0.317 & 0.267 & 0.050 & -0.027 & -0.098 \\
52 & 1-3-3-4 & One pair & 0.331 & 0.260 & 0.071 & -0.022 & -0.070 \\
53 & 1-1-2-4 & One pair & 0.359 & 0.270 & 0.089 & -0.015 & -0.085 \\
54 & 2-2-4-5 & One pair & 0.365 & 0.338 & 0.027 & -0.014 & -0.089 \\
55 & 2-2-2-4 & Triple + singleton & 0.345 & 0.296 & 0.050 & -0.018 & -0.100 \\
56 & 4-5-5-5 & Triple + singleton & 0.503 & 0.442 & 0.061 & -0.009 & -- \\
57 & 1-1-1-2 & Triple + singleton & 0.366 & 0.243 & 0.123 & -0.016 & -0.087 \\
58 & 1-3-3-3 & Triple + singleton & 0.283 & 0.235 & 0.048 & -0.032 & -0.069 \\
59 & 1-5-5-5 & Triple + singleton & 0.435 & 0.397 & 0.038 & -0.010 & -0.044 \\
60 & 1-1-1-3 & Triple + singleton & 0.344 & 0.233 & 0.111 & -0.021 & -0.078 \\
61 & 3-3-4-4 & Two pairs & 0.416 & 0.290 & 0.126 & -0.023 & -- \\
62 & 3-4-4-4 & Triple + singleton & 0.499 & 0.321 & 0.178 & -0.017 & -- \\
63 & 1-4-4-4 & Triple + singleton & 0.462 & 0.320 & 0.142 & -0.010 & -0.073 \\
64 & 1-1-1-5 & Triple + singleton & 0.397 & 0.278 & 0.119 & -0.013 & -0.073 \\
65 & 1-1-3-5 & One pair & 0.334 & 0.279 & 0.055 & -0.020 & -0.063 \\
66 & 4-4-4-4 & All equal & 0.601 & 0.356 & 0.245 & -0.010 & -- \\
67 & 2-4-4-4 & Triple + singleton & 0.479 & 0.335 & 0.145 & -0.011 & -0.096 \\
68 & 2-2-2-5 & Triple + singleton & 0.315 & 0.318 & -0.003 & -0.019 & -0.092 \\
69 & 3-3-3-3 & All equal & 0.297 & 0.236 & 0.061 & -0.042 & -- \\
70 & 2-3-3-3 & Triple + singleton & 0.308 & 0.246 & 0.062 & -0.035 & -0.092 \\
\bottomrule
\end{longtable}

\end{landscape}

\subsection{Operating characteristics stratified by true OBD}

\begin{figure}[!htbp]
\centering
\includegraphics[width=\textwidth]{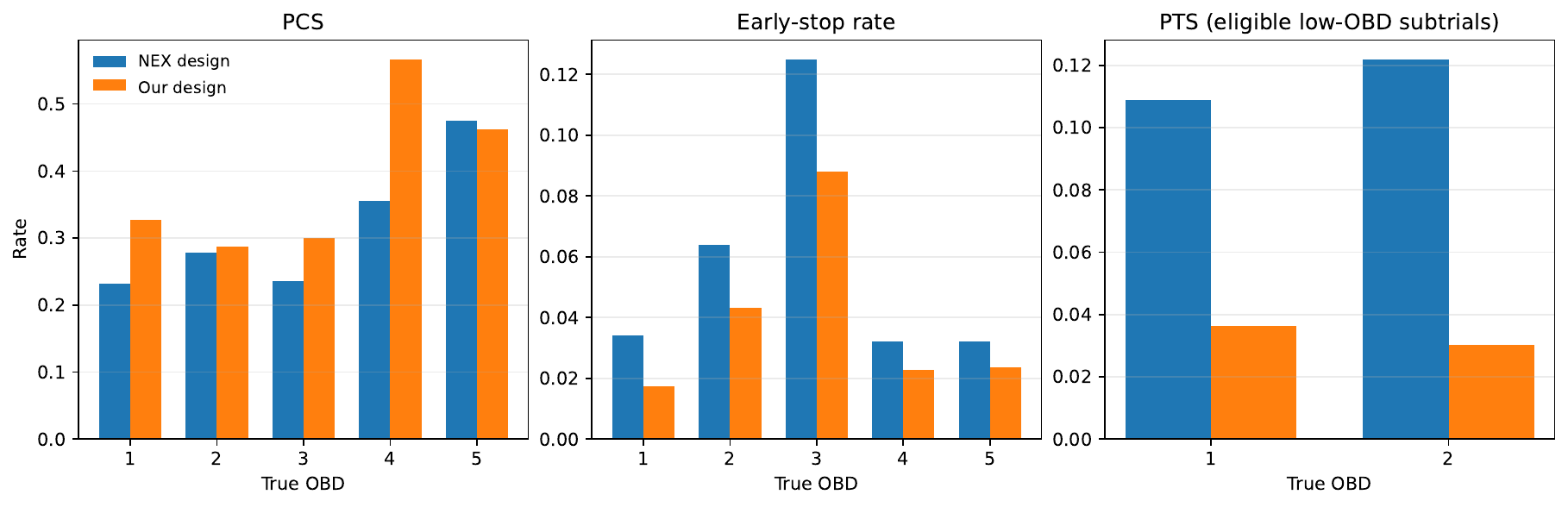}
\caption{Subtrial-level operating characteristics stratified by the true OBD. The PTS panel is restricted to low-OBD subtrials with true OBD 1 or 2.}
\label{fig:obd-stratified}
\end{figure}

Table~\ref{tab:obd-stratified} and Figure~\ref{fig:obd-stratified} summarise the average operating characteristics by true OBD location. The PTS panel is restricted to low-OBD subtrials in which overly toxic dose selection is applicable.

\begin{table}[!htbp]
\centering
\caption{Averaged operating characteristics stratified by the true OBD.}
\label{tab:obd-stratified}
\resizebox{\textwidth}{!}{%
\begin{tabular}{rrrrrrrr}
\toprule
True OBD & No. subtrials & PCS$_{FEX}$ & PCS$_{NEX}$ & PCS gain & Drop$_{FEX}$ & Drop$_{NEX}$ & Drop gain\\
\midrule
1 & 56 & 0.327 & 0.232 & 0.095 & 0.017 & 0.034 & -0.017 \\
2 & 56 & 0.288 & 0.278 & 0.010 & 0.043 & 0.064 & -0.021 \\
3 & 56 & 0.300 & 0.236 & 0.064 & 0.088 & 0.125 & -0.037 \\
4 & 56 & 0.567 & 0.356 & 0.211 & 0.023 & 0.032 & -0.009 \\
5 & 56 & 0.462 & 0.475 & -0.013 & 0.024 & 0.032 & -0.008 \\
\bottomrule
\end{tabular}%
}
\end{table}

\subsection{Over-borrowing-sensitive scenarios}

Figure~S1 displays the final selection distribution of the DF-EXNEX design in scenarios with negative or near-zero geometric mean PCS gain. Red boxes mark the selection probability for true OBD. The figure illustrates that small losses are not caused by random instability, but by coherent shifts toward nearby or intermediate doses. In scenarios combining low and high OBDs, the DF-EXNEX design can place substantial mass on dose 3 or 4 even when the true OBD is 2 or 5.

\begin{figure}[!htbp]
\centering
\includegraphics[width=0.95\textwidth]{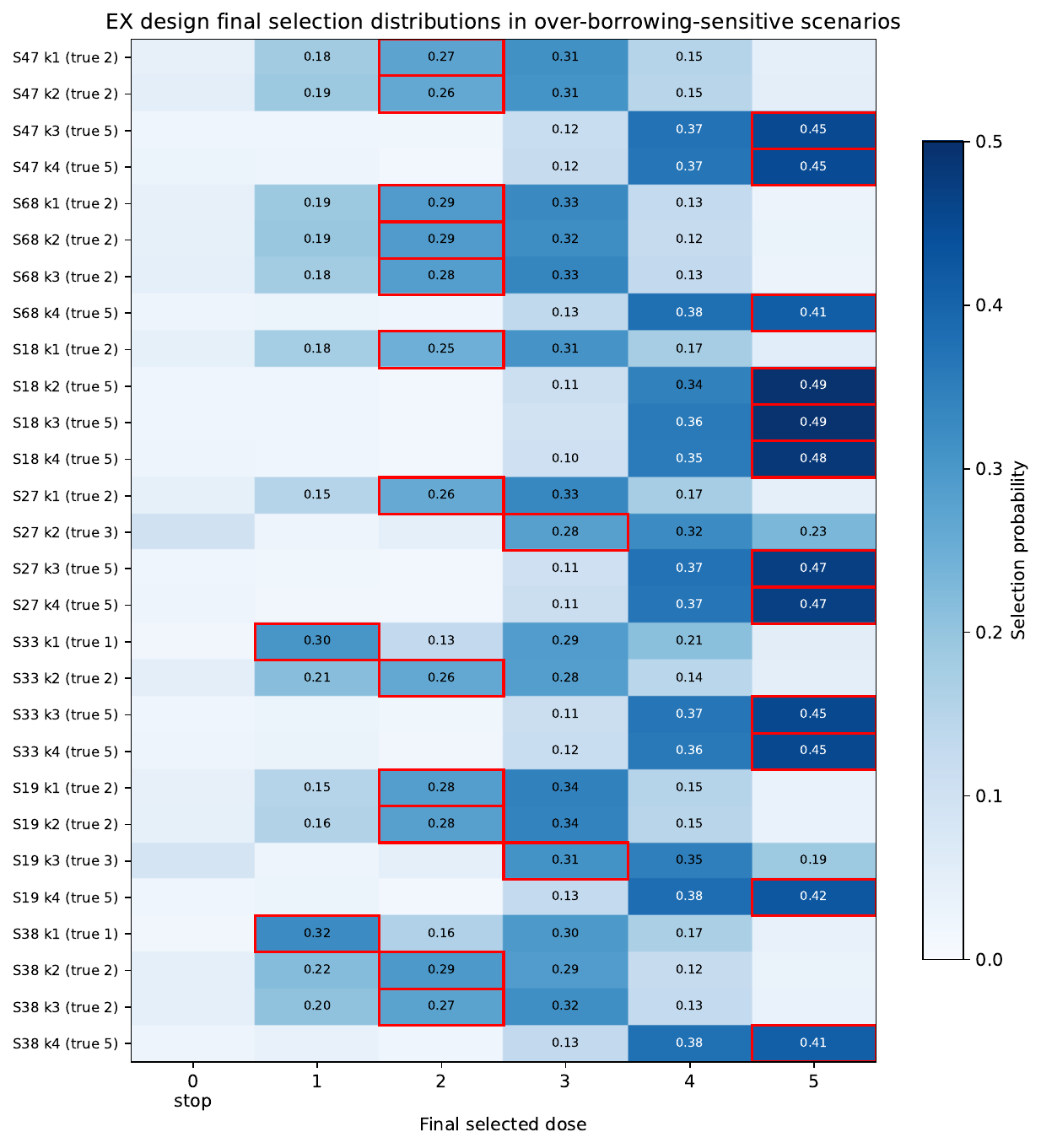}
\caption{DF-EXNEX design final selection distributions in selected borrowing sensitive scenarios. Rows are scenario-subtrial combinations: for example S47k1 means subtrial 1 in scenario 47 as shown in Table~\ref{tab:full-scenario-results}; columns are final selected doses, where dose 0 denotes early stopping. Red boxes mark the true OBD.}
\label{fig:overborrowing-heatmap}
\end{figure}

\begin{comment}
\input{tables/overborrowing_diagnostics.tex}

\subsection{Reference diagnostics from the complementary analyses}

The complementary 8-approach analyses were not used as the primary numerical source for the manuscript. They are nevertheless helpful as reference diagnostics. Table~S4 summarises performance by true-OBD structure, and Table~S5 summarises performance by exact and near-borrowability scores. These summaries support the main conclusion that borrowing gains increase with similarity, while the few losses are concentrated in structures where borrowing may act across separated OBD locations.

\input{tables/structure_diagnostics.tex}

\input{tables/borrowability_diagnostics.tex}
\end{comment}

%\bibliographystyle{plainnat}
%\bibliography{references}